\documentclass[12pt,preprint]{aastex}

\shorttitle{Nearby Spiral Globular Cluster Systems II}
\shortauthors{Nantais et al.}

\begin{document}

\title{Nearby Spiral Galaxy Globular Cluster Systems II: Globular Cluster Metallicities in NGC 300$^{1,2,3}$}

\author{Julie B. Nantais and John P. Huchra}
\affil{Harvard-Smithsonian Center for Astrophysics}
\affil{60 Garden Street, Cambridge, MA 02138}
\author{Pauline Barmby}
\affil{University of Western Ontario}
\affil{London, ON N6A 3K7, Canada}
\and
\author{Knut A. G. Olsen}
\affil{Cerro-Tololo Inter-American Observatory, National Optical Astronomy Observatory}
\affil{Casilla 603, La Serena, Chile}

\footnotetext[1]{Data for this project were obtained at the Baade 6.5 m telescope, Las Campanas Observatory, Chile.}
\footnotetext[2]{This study uses observations from the Hubble Space Telescope obtained at the Space Telescope Science Institute, operated by the Association of Universities for Research in Astronomy, Inc., under NASA contract NAS 5-26555.  These observations are associated with Programs GO-9162, GO-9492, and GO-10915.}\footnotetext[3]{This publication makes use of data products from the Two Micron All Sky Survey, which is a joint project of the University of Massachusetts and the Infrared Processing and Analysis Center/California Institute of Technology, funded by the National Aeronautics and Space Administration and the National Science Foundation.}

\begin{abstract}
We present new metallicity estimates for globular cluster (GC) candidates in the Sd spiral NGC 300, one of the nearest spiral galaxies outside the Local Group. We have obtained optical spectroscopy for 44 Sculptor Group GC candidates with the Boller and Chivens (B\&C) spectrograph on the Baade Telescope at Las Campanas Observatory.  There are 2 GCs in NGC 253 and 12 objects in NGC 300 with globular-cluster-like spectral features, 9 of which have radial velocities above 0 km s$^{-1}$.  The remaining three, due to their radial velocities being below the expected 95\% confidence limit for velocities of NGC 300 halo objects, are flagged as possible foreground stars.  The non-clusterlike candidates included 13 stars, 15 galaxies, and an H{\small{II}} region.  One GC, four galaxies, two stars, and the H{\small{II}} region from our sample were identified in archival Hubble Space Telescope images.  For the GCs, we measure spectral indices and estimate metallicities using an empirical calibration based on Milky Way GCs.  The GCs of NGC 300 appear similar to those of the Milky Way.  Excluding possible stars and including clusters from the literature, the GC system (GCS) has a velocity dispersion of 68 km s$^{-1}$, and has no clear evidence of rotation.  The mean metallicity for our full cluster sample plus one literature object is [Fe/H] = $-0.94$, lying above the relationship between mean GC metallicity and overall galaxy luminosity.  Excluding the three low-velocity candidates, we obtain a mean [Fe/H] = $-0.98$, still higher than expected, raising the possibility of significant foreground star contamination even in this sample.  Visual confirmation of genuine GCs using high-resolution space-based imagery could greatly reduce the potential problem of interlopers in small samples of GCSs in low-radial-velocity galaxies.
 
\end{abstract}

\keywords{galaxies: star clusters: general---galaxies: individual (NGC 300)---galaxies: spiral}

\section{Introduction}
Globular clusters (GCs), relics of some of the earliest and/or most violent phases of star and galaxy formation, can be analyzed to understand how various types of galaxies formed, as described in the review by \citet{bro06}.  Two subpopulations of GCs, metal-rich (red) and metal-poor (blue), exist in most early-type galaxies.  Both are thought to be old, but the red metal-rich GCs are thought to be slightly younger \citep{lar01,kun98,lee98}.  Large spiral galaxies, including the Milky Way \citep{zin85} and Andromeda \citep{bar00}, also have metal-rich and metal-poor GC subpopulations.  In smaller, later-type galaxies, however, often only a metal-poor population is seen \citep{cha04}.  

The three basic models for bimodal GC system (GCS) formation in early type galaxies are dwarf galaxy accretion \citep{cot98}, in situ formation \citep{for97}, and gas-rich mergers \citep{ash92}.  In the last scenario, the original spiral galaxies --- likely late-type spirals --- would provide the entire metal-poor GC population, and the metal-rich GC population would form in the merger.  Some evidence for GCs resulting from every one of these processes exists, but the properties of both blue and red GC populations have been found to scale with the mass of the host galaxy \citep{str04}, indicating that the accretion model alone cannot account for the majority of blue GCs in bright galaxies.  In any case, at high redshift and with a hierarchical merging scenario as the main process of galaxy formation, the distinction between these models fades.

The nearby Sculptor Group (or filament) is home to several late-type galaxies (Hubble types Sc--Im), including NGC 253 (the largest), NGC 300, NGC 55, and NGC 45.  At a distance of 1.9 Mpc \citep{gie04}, NGC 300 is one of the nearest spiral galaxies outside of the Local Group.  As a nearby example of a late-type, midsize spiral galaxy, it is especially useful for understanding the sparse and thus relatively poorly studied GCSs of late-type galaxies.  \citet{kim02} studied the star clusters in NGC 300 photometrically and identified 17 objects as GC candidates based on their size, shape, and color.  \citet{ols04} performed photometry on fields in six Sculptor Group galaxies and identified GC candidates via shape and color, and then obtained spectra for 19 GCs.  Six of these spectroscopically confirmed clusters had high enough signal-to-noise spectra for metallicity determination via Lick/IDS indices.  Seven of the GCs Olsen et al. spectroscopically confirmed were in NGC 300, and two had spectroscopically determined metallicities.

In this paper we present new metallicities derived from spectroscopy of GCs in Sculptor Group galaxies, primarily NGC 300, using \citet{kim02} and \citet{ols04} as reference catalogs.  We compare the metallicities of NGC 300 GCs to those of M31 and Milky Way clusters, and also determine NGC 300's place on the galaxy luminosity-GCS metallicity relation of \citet{bro91}.

\section{Data reduction and cluster selection}
We obtained spectra of 44 Sculptor Group cluster candidates with the Boller \& Chivers (B\&C) spectrograph on the Baade telescope from 2002 November 6 to 2002 November 9.  We chose to observe all 17 GC candidates from \citet{kim02}, plus 25 \citet{ols04} NGC 300 GC candidates with no previous spectra (except for NGC 300ax) observed in order of increasing magnitude from brightest to faintest.  We also observed the two brightest unconfirmed Olsen et al. GC candidates in NGC 253.  Nineteen of our 27 Olsen et al. cluster candidates are among the 38 Sculptor Group GC candidates for which we also obtained {\it JHK} photometry with the PANIC camera on the Baade 6.5 m telescope \citep[Paper I]{nan06}.  Figure 1 shows the locations of the NGC 300 objects we observed with respect to a schematic of the NGC 300 disk.  

The advantages of using the long-slit B\&C spectrograph rather than a fiber spectrograph are good coverage in blue optical wavelengths and better sky subtraction (due to the sky background being determined at the same time and place as the object exposure).  Also, the sky at Las Campanas has significantly weaker telluric emission lines than less isolated telescope locations.  

Our spectra were taken with the 600 l mm$^{-1}$ grating blazed at 5000 \AA.  The wavelength range is about 3700-6860 \AA.  Peak count rates occurred between 5000 and 5500 {\AA} for objects with typical GC colors.  The dispersion of the spectra is $\sim$1.6 {\AA} pixel$^{-1}$, and the resolution of the spectra is about 5 \AA.  The readout noise is 3.1 e$^-$.  Total exposure times were as low as 180 s for very bright objects and as high as 3600 s for faint objects, with 600-2400 s exposure times for most objects.  Exposure times under 600 s were usually done as single exposures, while longer exposures were usually broken up into 2-4 exposures of 600-900 s each.  Seeing ranged from 0.6-1.9$\arcsec$ over the course of obesrvations, with 1$\arcsec$ being typical, and the slit width used was 1$\arcsec$.  The median total signal-to-nose ratio (all wavelengths combined, $\sim$3700-6875 {\AA}) was about 20 per pixel for foreground stars and GC candidates, with values ranging from 4 to over 100.  If background galaxies, easily distinguished from these objects by their radial velocities, are included, the mean S/N per pixel is about 15.

Bias subtraction (with an averaged bias image), dark count correction, and flatfielding were done with IRAF's ccdproc task, and spectra were extracted using IRAF's apall task.  Other tasks in the specred and onedspec packages were used to combine spectra, calibrate wavelengths, flux-calibrate the spectra using standard star observations taken each night, and eliminate pollutants such as cosmic rays and improperly-subtracted telluric lines.  

Velocities were determined with the xcsao task in the rvsao package \citep{kur98} in IRAF using a list of standard SAO templates useful for the expected features of GCs, background galaxies, and foreground stars.  Included are four templates derived from M31 GCs (m31\_a\_temp, m31\_f\_temp, m31\_k\_temp, and fglotemp), three galaxy absorption templates (fm32temp, habtemp90, and fn4486btemp), two composite stellar absorption templates (fabtemp97 and fallstars), a synthetic H{\small{II}} region/galaxy emission template (hemtemp0.0), and a synthetic Ca H\&K absorption template (hkabstemp). For each object, we list a heliocentric velocity determined from the template with the highest R value.

Objects labeled as clusters usually had spectra similar to G and early K giants.  Mid-type K stars ($\sim$ K3-K5) typically differed visibly from GC candidates in having a redder spectral energy distribution and strong Mg and Ca42 features.  In order to provide a systematic, quantitative classification scheme to remove K dwarfs from the pool of GC candidates, we focused on the strong Ca42 feature and compared it to two other spectral features, G43 and H$\delta$ (using the H$\delta$A index definition in \citet{tra98}).  The difference between the Ca42 and H$\delta$ indices was used by \citet{per95} to distinguish between K dwarfs and GC candidates.  In our own samples, the Ca42 index also appeared strong in comparison to the G43 index in objects that had the spectral features of mid-K-type stars.  To quantify the differences in these features between GCs and K dwarfs, we measured the Ca42, H$\delta$A, and G43 indices in the 41 \citet{sch05} Milky Way GC spectra and in 16 K dwarfs from the Jones Coud\'eFeed spectral library described in \citet{lei96}.  The K dwarfs range in type from K0 to K8.  Figure 2 shows histograms of Ca42-G43 and Ca42-H$\delta$A ratios for the Milky Way GCs and Jones Coud\'eFeed K dwarfs respectively.  There is some overlap between the distributions of Ca42-G43 for GCs and K dwarfs, particularly among early K-type stars (K0-K2), but this index appears useful for screening out mid-K-type stars in our sample.  The Ca42-H$\delta$A ratio seems more effective in discriminating between GCs and K dwarfs, with the only K dwarf overlapping the MW GC range of values being a K8 star, which in our NGC 300 sample would be easily distinguishable from a GC.  The Milky Way GCs had a mean Ca42-G43 value of -0.057 with a $\sigma$ of 0.024 and a mean Ca42-H$\delta$A value of 0.016 with a $\sigma$ of 0.075.

For an object to be considered a GC candidate, we insisted that it have both Ca42-G43 $\leq$ 0.05 (the value for NGC 300-05 which, despite having a Ca42-G43 ratio more than 3$\sigma$ above the MW GCs, has been visually confirmed as a GC via HST archive images) and Ca42-H$\delta$A $\leq$ 0.241 ($3\sigma$ above the Milky Way GC mean).  This analysis excludes NGC 300cr and NGC 300df on the basis of Ca42-H$\delta$A but not Ca42-G43 (which one might expect for K0-K2 dwarfs), and NGC 300ax, previously classified by \citet{ols04} as a GC, on the basis of Ca42-G43 but not Ca42-H$\delta$.

Radial velocity can provide limited additional information on the likelihood of an object being a star or a GC.  Since the heliocentric radial velocity of NGC 300 is fairly low --- +142 $\pm$ 4 km s$^{-1}$ \citep{rc3} --- it is impossible to eliminate all stars and keep all GCs in the sample if a strict velocity cut is used to exclude candidates.  Hence, we cannot automatically label objects with clusterlike spectra but very low radial velocities as definite stars.  The velocity dispersion in NGC 300's halo is estimated by \citet{car85} to be 60 km s$^{-1}$.  If the Milky Way halo were non-rotating, the average expected radial velocity of a Galactic star in the direction of NGC 300 would be about +40 km s$^{-1}$, and the velocity dispersion of the Milky Way halo is about 130 km s$^{-1}$ (as viewed from the center of the Milky Way).  So while there's a 95\% chance that an NGC 300 halo object will have a radial velocity greater than 20 km s$^{-1}$, Milky Way halo stars can very easily have velocities similar to those of genuine NGC 300 GCs.  This means that most objects with negative radial velocities will be Milky Way stars, but those with positive radial velocities could easily be either foreground stars or NGC 300 clusters, and only spectral features or high-resolution imaging can distinguish them.  On the basis of radial velocity, we flag three objects with spectral features similar to GC candidates - NGC 300cm, NGC 300co, and NGC 300db - as possible stars.  All have radial velocities less than 0 km s$^{-1}$.

Figure 3 shows the correlations between our two indicators of star vs. GC status, Ca42-G43 and Ca42-H$\delta$A, and the radial velocities of foreground stars and GC candidates in NGC 300.  Both trends are high in scatter, demonstrating the difficulty in measuring star vs. GC status on the basis of radial velocity.  The correlation between Ca42-H$\delta$A and radial velocity is somewhat stronger, probably due to its greater sensitivity to K0-K2 dwarfs compared to Ca42-G43.

The best-fit velocity template also hints at whether an object is likely to be a star or a cluster.  The templates we used included several templates developed using intermediate-age and old stellar populations (m31\_a\_temp, m31\_f\_temp, m31\_k\_temp, fglotemp, and fm32temp), a couple templates suited for field stars (fallstars, fabtemp97), and a couple synthetic templates (hemtemp0.0 for H{\small{II}} emission and hkabstemp for Ca {\small{II}} H \& K absorption). Most GCs were best fit by the M31 F and K templates, while many foreground dwarfs were best fit by the stellar absorption templates.

Figure 4 shows sample spectra of an old GC, a typical foreground K star, and a background galaxy.  The region shown in the plot covers wavelengths from 3700 to 5300{\AA}, containing key features such as CNB, Ca{\small{I}} 4227, G4300, H$\beta$, Mgb, and several Fe features.  One can see the K star's strong Ca{\small{I}} 4227 and Mgb features, compared to the relatively weak Ca{\small{I}} 4227 and Mgb and strong G4300 of the GC.  One can also see that in the foreground star, the continuum on the blue side of the Mg2 feature tends to be notably fainter than on the red side.  The galaxy is recognizable by its visibly redshifted Ca{\small{II}} H\&K lines.

\section{Discovering Clusters}

The names, heliocentric cross-correlation velocities, and object types (background galaxy, foreground star, cluster candidate) are shown in Table 1.  The objects with number designations 1-17 are from Kim et al. 2002; the rest are from Olsen et al. 2004.  Of the 17 Kim et al. candidates, only five --- objects 2, 3, 4, 5, and 12 --- appear to be old GCs.  Note that object 3 is also NGC 300a in Olsen et al.'s catalog, and has been proven a GC and spectroscopically evaluated by Olsen and collaborators.  Most of the remaining Kim et al. objects were background galaxies, except for one H{\small{II}} region (Object 6) and three probable NGC 300 stars (NGC 300-09, M-type; NGC 300-10, F-type; and NGC 300-15, A-type).

Since GCs in NGC 300 can easily be resolved with HST, we searched the HST ACS and WFPC2 archives for images of our NGC 300 objects.  The images were from programs GO-8584, GO-9162, GO-9492, and GO-10915.  We found eight Kim et al. objects in the archival HST images: Objects 1, 5, 6, 7, 8, 9, 10, and 11.  These eight objects, imaged in F555W for the ACS images of Objects 1, 5, 6, 7, and 11 and F606W for the WFPC2 images of Objects 8, 9, and 10, are shown in Figure 5.  Generally, their appearances match what we expect based on their spectra.  Objects 1, 7, and 11 are spiral galaxies; Object 8 is an early-type galaxy; Object 6 is an H{\small{II}} complex; Object 5 appears to be a genuine GC; and Objects 9 and 10, initially considered as possible open clusters, appear pointlike but non-saturated and without diffraction spikes.  Their \citet{kim02} V magnitudes are 19.73 for NGC 300-09 (which has an M-type spectrum) and 19.12 for NGC 300-10 (which has an F-type spectrum), consistent with supergiant stars in NGC 300.  They both have many fainter stars surrounding them, so may also be open clusters or stellar blends.

Kim et al. selected their GC candidates by using a color cut (0.3 $<$ $B-V$ $<$ 2.0) and morphological considerations, including visual inspection of the shapes and brightness profiles.  They assigned three ``classes'' to the quality of GC candidates, Class 1 being most likely to be GCs and Class 3 being least likely.  Interestingly, Class 2 yielded the highest fraction of objects that turned out to be GCs: 3 out of 7 had GC spectra, as compared to 1 out of 4 for Class 1 and 1 out of 6 for Class 3.  Kim et al. reported using morphological considerations to determine the probability of an object being a GC.  If they required that the clusters be resolved from the ground to be Class I objects, they may have actually shifted the balance out of favor of true GCs and in favor of galaxies, since a typical GC at the distance of NGC 300 is close to 1$\arcsec$ in diameter, which is the limit imposed by typical ground-based seeing.

Of the Olsen et al. candidates, 11 objects appeared to be stars, including NGC 300ax, which Olsen et al. had identified and analyzed as a GC.  Its spectrum has wide spectral lines consistent with a dwarf of late K to early M type: a wide but shallow Mgb (5180 {\AA}) feature and a wide and deep NaD (5895 {\AA}) feature, as well as having a strong Ca{\small{I}} 4227 feature as compared to the G4300 feature much like other K dwarfs.  In addition to this, 3 more GC candidates with Ca42-G43 $\leq$ 0.05 and Ca42-H$\delta$A $\leq$ 0.241 have been flagged as possible stars on the basis of radial velocities less than 0 km s$^{-1}$.

\section{Velocity Dispersion, Rotation, and Estimated Mass}
Our GC velocities --- 9 highly probable clusters and 3 low-velocity possible clusters --- can be combined with seven previously observed GC velocities, five from \citet{ols04} and two from unpublished data contributed by one of us (K. Olsen), to obtain a GC velocity dispersion and simple mass estimate for NGC 300.  The velocities of the seven previously observed clusters are listed in Table 2.  Figure 6 shows the velocity histogram of all GC candidates, both previously observed by Olsen et al. and observed in this paper, including those with V$<$0.

If we include all highly probable and possible cluster candidates --- 19 objects in all --- we have a mean velocity of 79 km s$^{-1}$ and a velocity dispersion, corrected for the mean uncertainty of the velocities, of 92 km s$^{-1}$.  If we exclude the three V$<$0 objects, we have 16 objects with a mean velocity of 107 km s$^{-1}$ and an uncertainty-corrected velocity dispersion of 68 km s$^{-1}$.  This latter dispersion is closer to the \citet{car85} theoretical velocity dispersion estimate of 60 km s$^{-1}$ than is the velocity dispersion of all 19 objects.

We can also investigate whether there is evidence for rotation among our GCs.  Figure 7 shows the velocity of cluster candidates as a function of position angle, along with a fit to the rotation of all candidates not labeled as possible stars and a representation of the \citet{puc90} H{\small{I}} rotation curve (with the velocity of 90 km s$^{-1}$ reduced by a factor of sin(i) with i = 42.3\degr).  We fit a function of the form 
\begin{equation}
V(\theta) = V_c + V_{pr}*sin(\theta - \theta_0)
\end{equation}
to the velocities and position angles of the 16 cluster candidates with V$>$0, presumed to be the best cluster candidates.  $V_c$ is the central velocity, $V_{pr}$ is the projected rotational velocity (the actual rotational velocity times sin(i)), $\theta$ is the position angle of the cluster in degrees measured east from north, and $\theta_0$ is the position angle of the rotation axis.  Using a nonlinear least-squares fit with weights equal to $1/\sigma_v^2$, fitted to all V$<$0 clusters, we found $V_c = (122 \pm 21)$ km s$^{-1}$, $V_{pr} = (91 \pm 32)$ km s$^{-1}$, and $\theta_0 = (27 \degr \pm 17\degr)$.  However, this rotation curve seems exceptionally strong and out of phase with the H{\small{I}} rotation, and the uncertainty-corrected dispersion of the GC velocities minus their predicted rotational velocities is actually considerably worse than our previous result (80 km s$^{-1}$).  An unweighted fit to the same points gives a similarly unsatisfactory result: $V_c = 122 \pm 22$ km s$^{-1}$, $V_{pr} = (42 \pm 34)$ km s$^{-1}$, and $\theta_0 = (164 \degr \pm 35 \degr)$, and still a worse velocity dispersion than without rotation (81 km s$^{-1}$).  We therefore conclude that there is no sign of rotation in our GC sample.

We use the Projected Mass Estimator \citep{hei85} to estimate the total mass of NGC 300, both using the means of the total and ``good'' samples and the mean velocity of NGC 300 itself.  The equation for the projected mass estimator is as follows:
\begin{equation}
M_{PM} = \frac{f_{pm}}{G(N-\alpha)} \sum_{i} V^2_{zi}R_{\bot i}.
\end{equation}

Here we use $f_{pm} = 32/\pi$ (for isotropic orbits, recommended by \citet{hei85} and $\alpha = 1.5$.  Using the published mean velocity of NGC 300 itself gives $M_{PM} = (1.8 \pm 0.8) \times 10^{11} M_{\odot}$ for the total sample, and $M_{PM} = (8.3 \pm 2.1) \times 10^{10} M_{\odot}$ for the V$>$0 subsample.  The total sample reaches a maximum projected radius of 12.1 kpc, and the V$>$0 sample reaches a maximum of 11.6 kpc.  Uncertainties were estimated using the bootstrap method with 1000 resamplings.  \citet{puc90}, using H{\small{I}} rotation curve data, calculate a mass of $2.4 \times 10^{10} M_{\odot}$ out to 10.6 kpc; \citet{rhe92} give mass-model-based estimates for the mass of NGC 300 ranging from $2.8-4.2 \times 10^{10} M_{\odot}$.  Overall, our masses run large but are not outside the realm of possibility.

\section{Metallicity Analysis}

 We measured 26 spectral indices,including the 12 indices calibrated in \citet{bro90} using the \citet{bro90} bandpass definitions and the diagnostic H$\delta$A index with the \citet{tra98} bandpass definition.  Eight of the Brodie and Huchra indices are mostly similar to the Lick/IDS bandpass definitions, differing slightly from those of \citet{wor94} and \citet{tra98}.  All spectral indices not in \citet{bro90} were measured using the bandpass definitions of \citet{tra98}.  All bandpasses were shifted to the uncorrected, geocentric radial velocities of our objects.  Our indices are not true Lick indices because our spectra have not been degraded to match the resolution of the Lick star spectra, and we did not observe any Lick standard stars to calibrate the fluxes and indices precisely.  However, we did observe a template galaxy (NGC 1052) that was also observed by \citet{bro90}, and found the overlapping indices to be similar to theirs.  All indices are given in magnitudes, as in Brodie and Huchra, for easy comparison with their M31 spectral indices and index-metallicity calibration methods.  Converting to equivalent width requires multiplication by the bandwidth of the central index band shifted to the object's velocity.  

We determined metallicities using a method based on the \citet{bro90} prescription, but with new index-vs.-metallicity calibrations determined from the 41 Milky Way spectra from \citet{sch05}, degraded to match the 5 \AA\ resolution of our spectra.  We determined linear regressions for metallicity (from \citet{har96} for each Milky Way GC) as a function of index strength for all our indices, as well as the ``range'' $R_I$ defined as in \citet{bro90}, and the $\sigma_m$ representing the scatter about the regression line.  We used the $R_I$ and $\sigma_m$ values to construct weights similar to those in \citet{bro90}:
\begin{equation}
W_I = \frac{R_I}{(\sigma_m^2+\sigma_p^2+\sigma_s^2)^{1/2}}
\end{equation}
with $\sigma_p$ being the photometric uncertainty in the entire index, the quadrature sum of the uncertainties in each band.  To calculate $\sigma_s$, the measure of our ability to measure the same index repeatedly for the same object over the course of different nights and conditions, we used spectral index measurements for the standard star LTT 9239, which we observed each night.  We modified the formula for the uncertainty in a band (continuum or feature value for an index) from \citet{bro90} to the following:
\begin{equation}
snr = 1/\sigma_B = O/(O+S+R^2)^{1/2},
\end{equation}
where $O$ is the number of object counts per pixel, $S$ is the number of sky counts per pixel, and $R$ is the read noise.  The weighted average and uncertainty in the weighted average were determined as in \citet{bro90}.

Our choice of indices to use in the weighted average was determined by considering the correlation coefficient in each Milky Way linear fit, the signal-to-noise ratio in the NGC 300 objects, and the degree of scatter in NGC 300 index-index relations.  The final set of indices - Mg2, Fe5270, Fe5335, Ca4227, G4300, $\delta$, Ca4455, and CN2 - all had relatively high correlation coefficients in the Milky Way index-metallicity relations and relatively high signal-to-noise in typical NGC 300 clusters.

Table 3 lists measurements and uncertainties of the 12 indices calibrated in Brodie and Huchra and the H$\delta$A index used to help screen out foreground stars, and Table 4 gives the measurements and uncertainties of the remaining 13 indices.  Table 5 lists the linear relationships between index and metallicity for the 8 indices we chose along with $\sigma_s$ measurements based on the standard star LTT 9239, and Table 6 lists derived metallicities for our clusters, along with five spectroscopically determined metallicities for Sculptor Group GCs from \citet{ols04}.  The highest $\sigma_s$ for spectral features used to measure metallicity was 0.095 (the $\delta$ index, in the region with the lowest signal-to-noise ratio and spanning a broad range of wavelengths) and the lowest was 0.001 (Fe5270, a narrow feature in the region with the highest signal-to-noise ratio).

Figures 8-12 compare various spectral indices to those of Andromeda Galaxy GCs \citep{hbk91,bar00} and the Schiavon et al. Milky Way GCs.  Figure 8 shows that the CNR index lies below typical values for Andromeda and Milky Way clusters, while the CNB index, traditionally considered a more reliable indicator of actual CN content, looks about the same as the other glaaxies.  Figure 9 shows a close-up of the CNB region for averages of 10 Milky Way GCs and 10 NGC 300 GCs matched for Fe52 strength.  The CNB depths look fairly similar, although a tilt due to the high reddening of many of the MW GCs is visible.  Fe52 metallicities, as seen in Figure 10, look similar in all three galaxies.  The G-band (Figure 11) seems about 0.02 mag high compared to the Milky Way and Andromeda, although overall more similar to the Milky Way.  And finally, H$\beta$, the index that is sensitive to age and inversely correlated with metallicity, is shown in Figure 12.  NGC 300 GCs show a mean and distribution more similar to the Milky Way than to Andromeda, suggesting old clusters similar to those of the Milky Way.

 The deficit in CNR in our sample does not necessarily indicate a nitrogen deficiency in the GCs of NGC 300, especially given the apparently normal CNB.  Andromeda GCs had once been thought to have nitrogen excesses compared to Milky Way GCs \citep{bur84,bro91}, but this was later challenged when the metal-rich Milky Way GCs were taken into account \citep{puz02}.  Instead, Puzia et al. found that metal-rich GCs in both Andromeda and the Milky Way were nitrogen-enhanced compared to the general old stellar populations of galaxies, which they propose may be due to low-mass GC stars accreting carbon and/or nitrogen from the AGB winds of dying higher-mass stars in the cluster.  A deficiency in CNR in our sample, therefore, may suggest a high rate of foreground star contamination especially among metal-rich objects, since stars should lack nitrogen compared to metal-rich GCs of similar metallicity.   However, CNB is normal, and CNR is a weak feature that can easily be trumped by noise, so there may not be any cause for alarm.  Two of the four objects on the lower right of the CNR vs. Mg2 plot (Figure 8(b)), with Mg2 indices above +0.125 and lacking strong CNR, are among the low-velocity candidates.

Overall, the NGC 300 GCs seem similar to those of the Milky Way.

Figure 13 shows the metallicity distribution of the NGC 300 GCs as compared to the Milky Way, Andromeda, and M33, including NGC 300r from \citet{ols04}.  With all candidates included, the distribution appears unimodal and rather metal-rich.  With the three low-velocity objects excluded, the distribution appears to still be fairly metal-rich.  The NGC 300 cluster sample is too small and possibly too contaminated to definitively determine bimodality or a lack thereof.

An updated version of the \citet{bro91} relation between mean GCS metallicity and total galaxy luminosity was created using relatively recent published galaxy distance and spectroscopic GC metallicity information from the literature on several galaxies, along with contemporary 2MASS K and \citet{rc3} B magnitudes found on the NASA/IPAC Extragalactic Database (NED).  Table 7 summarizes the data, gathered from a wide variety of studies of GCSs using different selection criteria.  The relationship between the mean GCS metallicity and absolute B magnitude is shown in Figure 14, and the relationship between mean GCS metallicity and absolute 2MASS K magnitude is shown in Figure 15.  NGC 300, with its calculated mean GC metallicity of $-0.94 \pm 0.15$, lies above both the early-type and late-type galaxies on the plot.  The high metallicity calculated for NGC 300 could be due to most of the ``higher metallicity'' cluster candidates being foreground stars, as suggested above in the discussion of CNR weakness.  If the three clusters deemed to be possible foreground stars are eliminated, the mean metallicity of the remaining ten clusters (our clusters plus NGC 300r) is $-0.98 \pm 0.12$, which is still higher than expected for a galaxy the size of NGC 300.  Besides possible foreground star contamination, color selection in the candidate samples may have created a bias in favor of metal-rich, intrinsically red clusters that drastically increases their relative numbers in our tiny sample.  This is unlikely for the Kim et al. clusters, but the \citet{ols04} color cuts were found to be $\sim$0.2 mag redder than initially thought due to a calibration error.  It is also worth noting that the spectroscopic metallicity estimates used in our metallicity-luminosity relations do not distinguish among the $[\alpha/Fe]$ values that may differ among galaxy types.

\section{Summary/Conclusions}

The GCS of NGC 300 appears to have features similar to that of the Milky Way.  Its place on the metallicity-luminosity relation is consistent with the positive correlation between galaxy luminosity and mean GC metallicity, although the average metallicity of our NGC 300 cluster sample is higher than expected.  Eliminating all ambiguous objects from our cluster sample gives a mean metallicity slightly more consistent with the Milky Way and Andromeda.  The GCs, excluding those deemed possible stars, have a velocity dispersion of 68 km s$^{-1}$, and may be rotating, although the evidence does not strongly favor rotation.  The high mean metallicity of our cluster sample may be primarily a result of the difficulty in screening out stars without the ability to use radial velocity cuts.  The problem of mistaking foreground stars (and giant stars within the target galaxy) for GCs could easily be remedied with more space-based imaging of NGC 300.  GCs could then be identified visually or with more accurate size and shape determinations before spectroscopic study.

\acknowledgments

We would like to thank the LCO staff for helping us use the Magellan Telescopes and instruments.  This work has been supported by the Smithsonian Institution and Harvard College Observatory.  We also thank Ricardo Schiavon and collaborators for making their Milky Way spectra available to the public at http://www.noao.edu/ggclib/.

\clearpage

\begin{figure}
\plotone{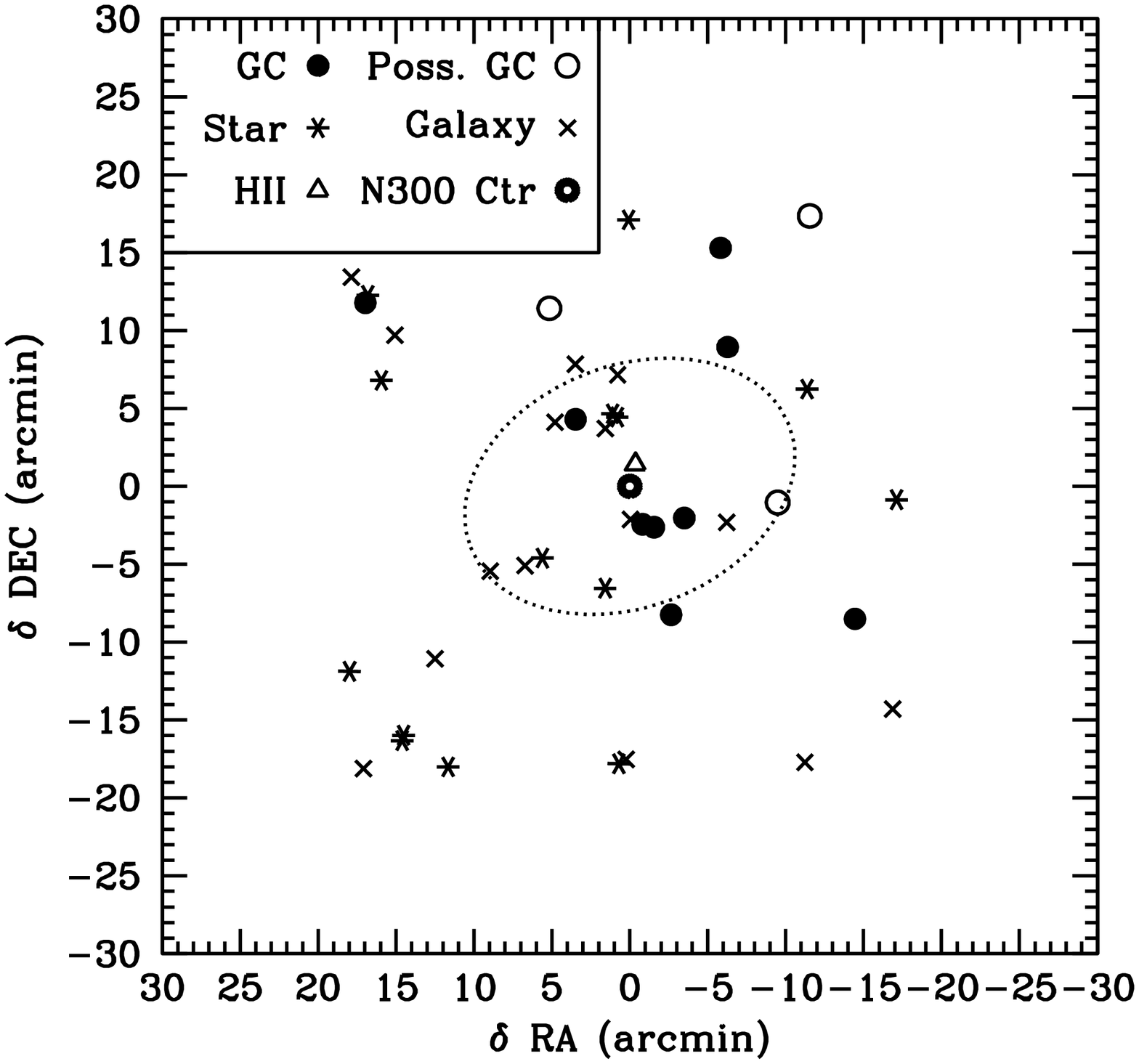}
\caption{Locations of various objects in NGC 300, including GCs, ``possible'' GCs (low velocity candidates), galaxies, stars, and H{\small{II}} regions.  The dotted line represents the disk of NGC 300.}
\end{figure}

\begin{figure}
\plottwo{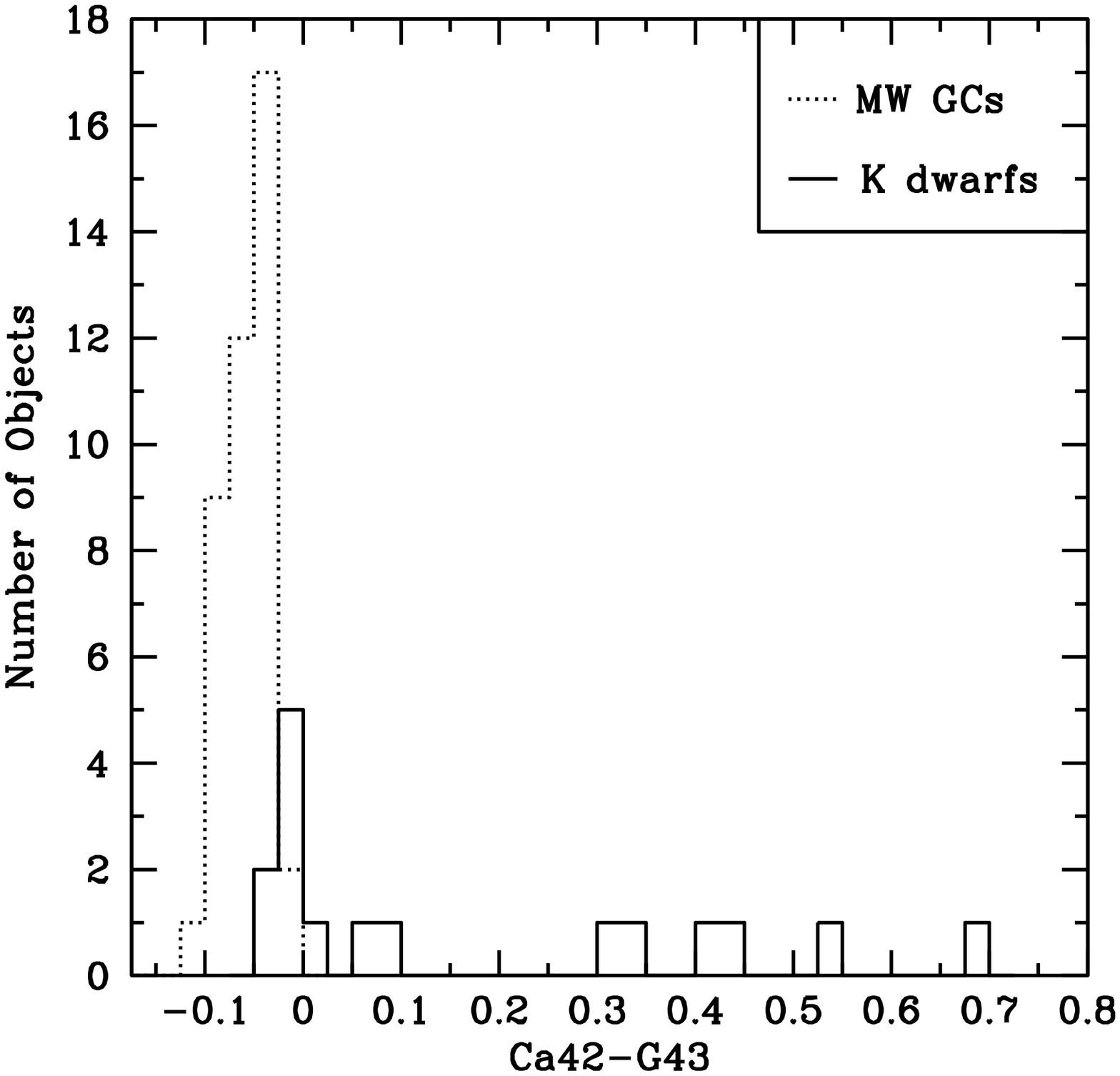}{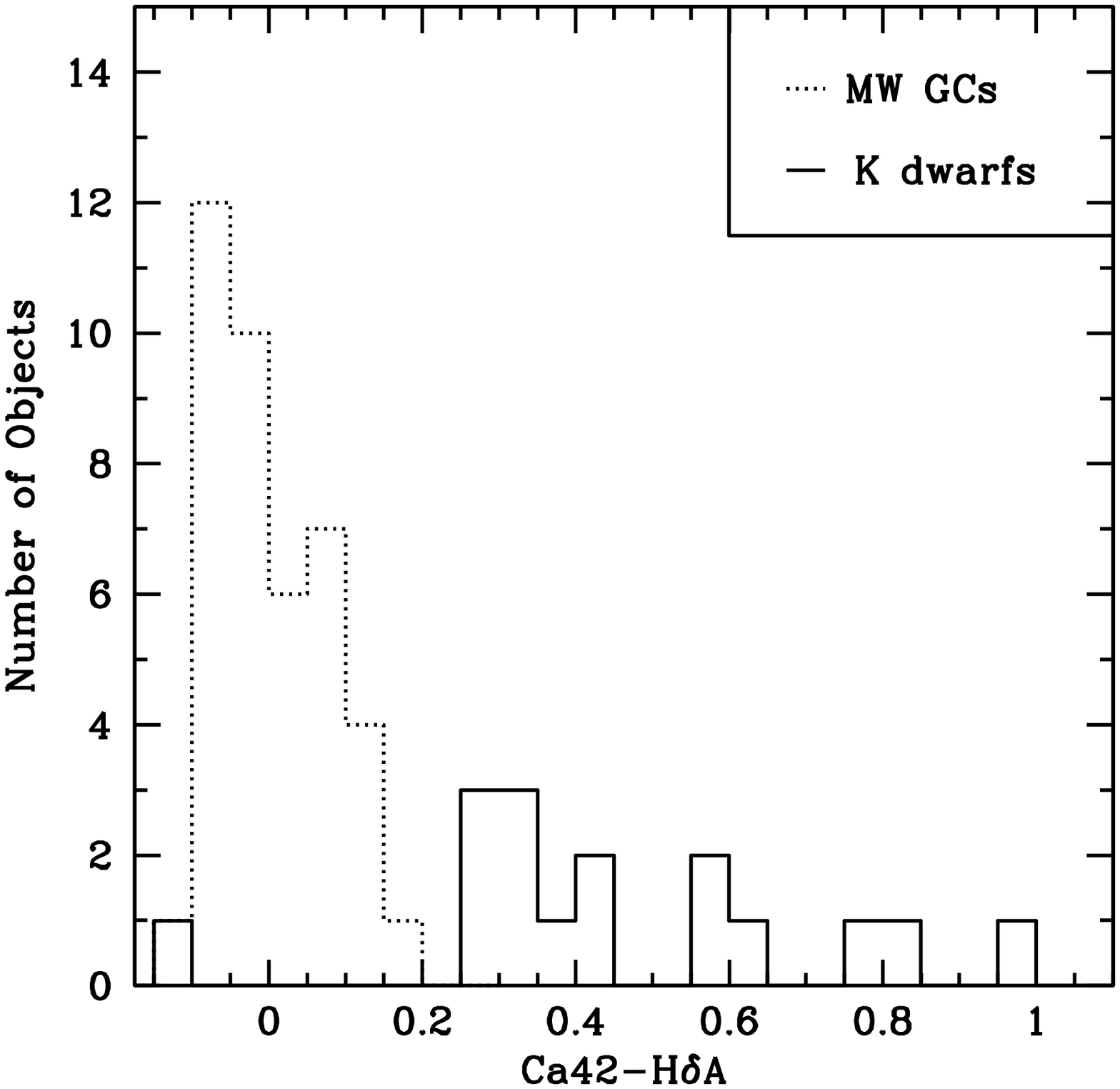}
\caption{Histograms of Ca42-H$\delta$A and Ca42-G43 index ratios for Milky Way GCs and Milky Way field K stars.}
\end{figure}

\begin{figure}
\plottwo{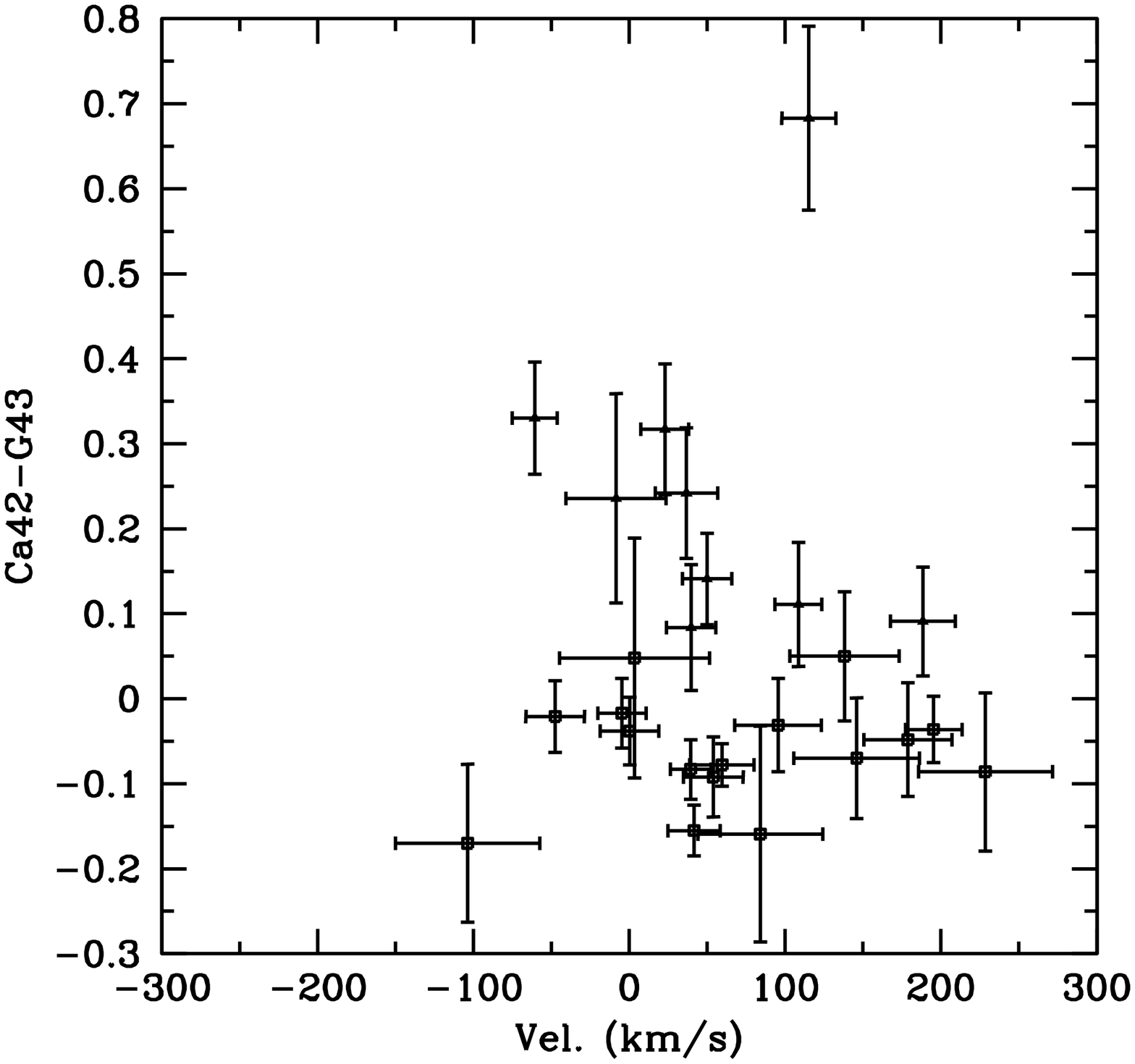}{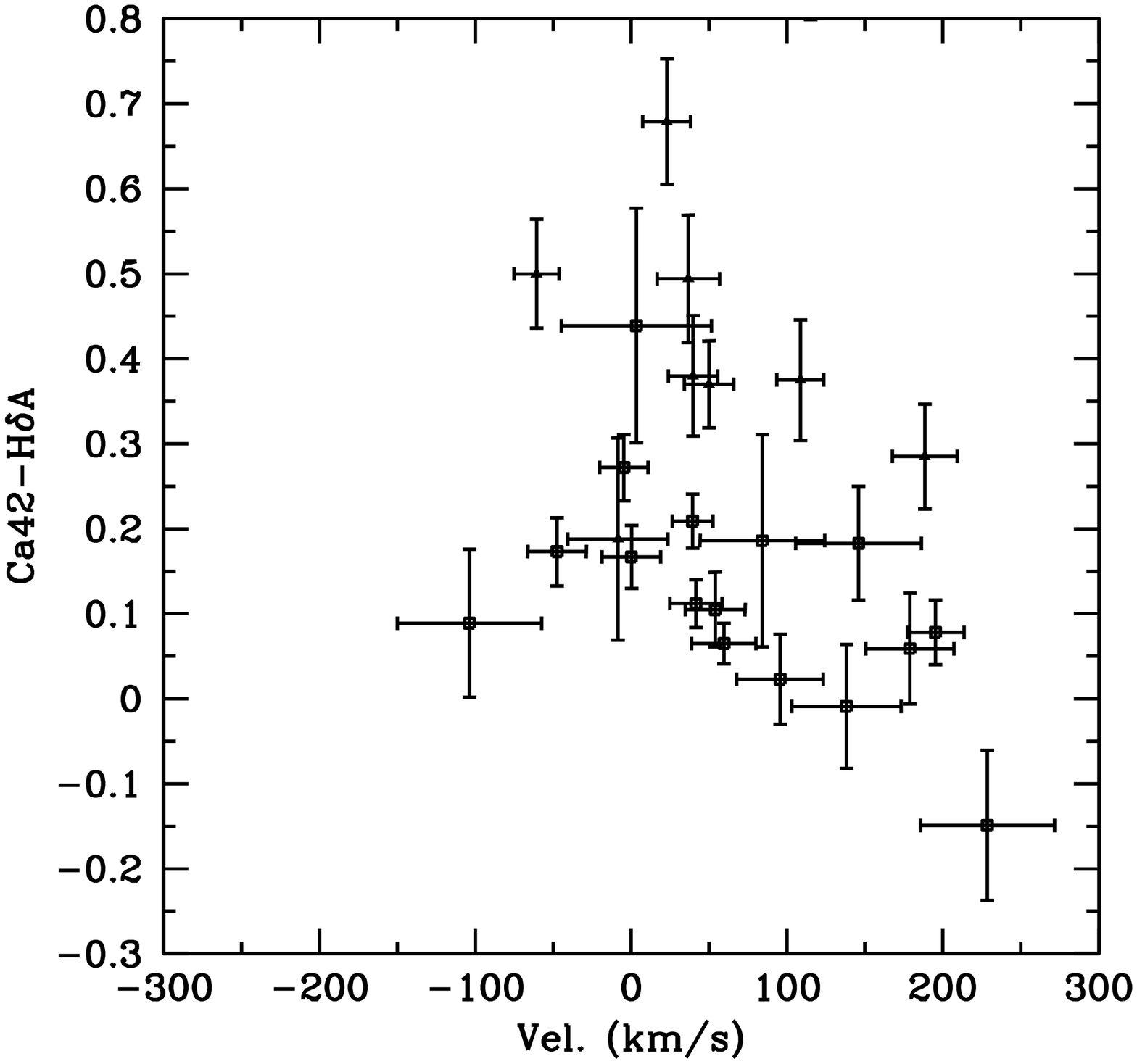}
\caption{Ca42-G43 (left) and Ca42-H$\delta$A (right) vs. radial velocity for all low-radial-velocity objects (foreground stars and GC candidates).}
\end{figure}

\begin{figure}
\plotone{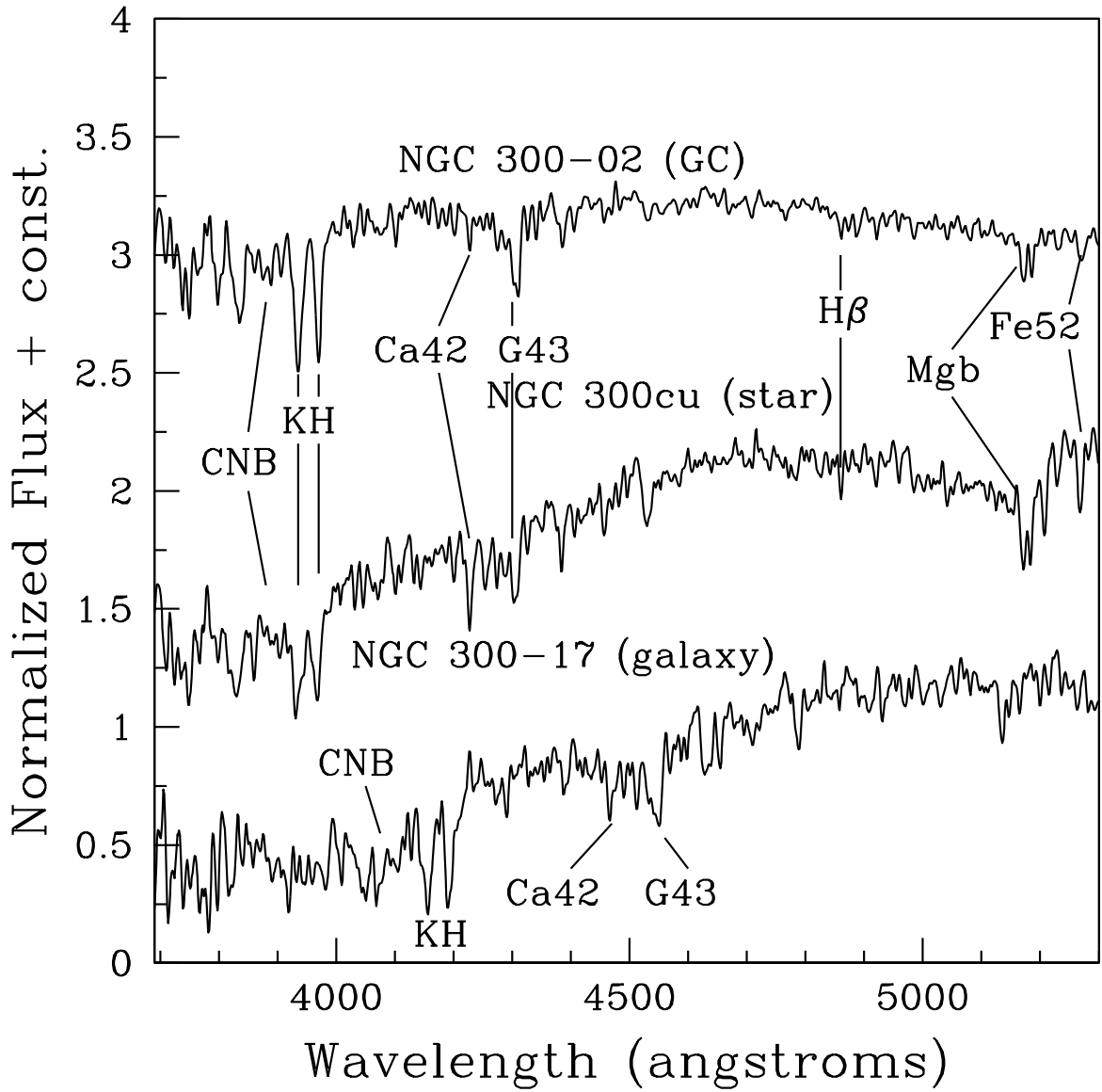}
\caption{Sample spectra of a GC (top), a foreground star (middle), and a background galaxy (bottom), normalized and shown on the same scale.}
\end{figure}

\begin{figure}
\plotone{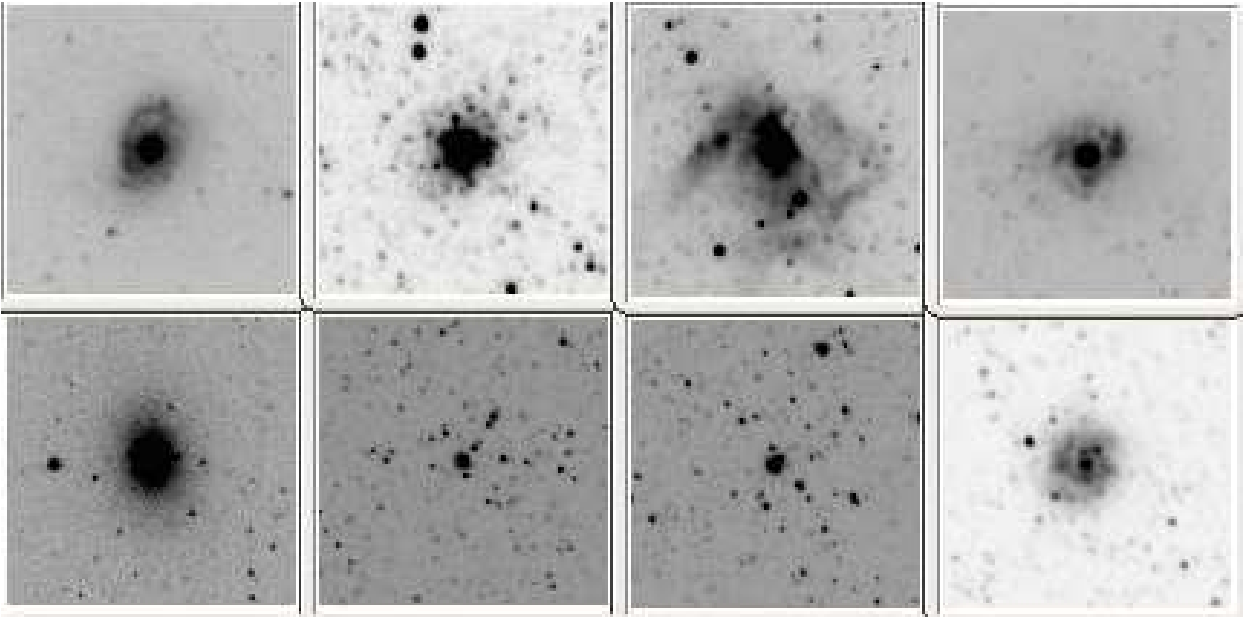}
\caption{NGC 300 objects from the Kim et al. catalog found in the HST archives.  On the top row, left to right, are Kim et al. Objects 1, 5, 6, and 7; on the bottom row, left to right, are Kim et al. objects 8, 9, 10, and 11.  Objects 1, 5, 6, 7, and 11 are imaged in F555W by ACS/WFC and Objects 8, 9, and 10 are imaged in F606W by WFPC2.}
\end{figure}

\begin{figure}
\plotone{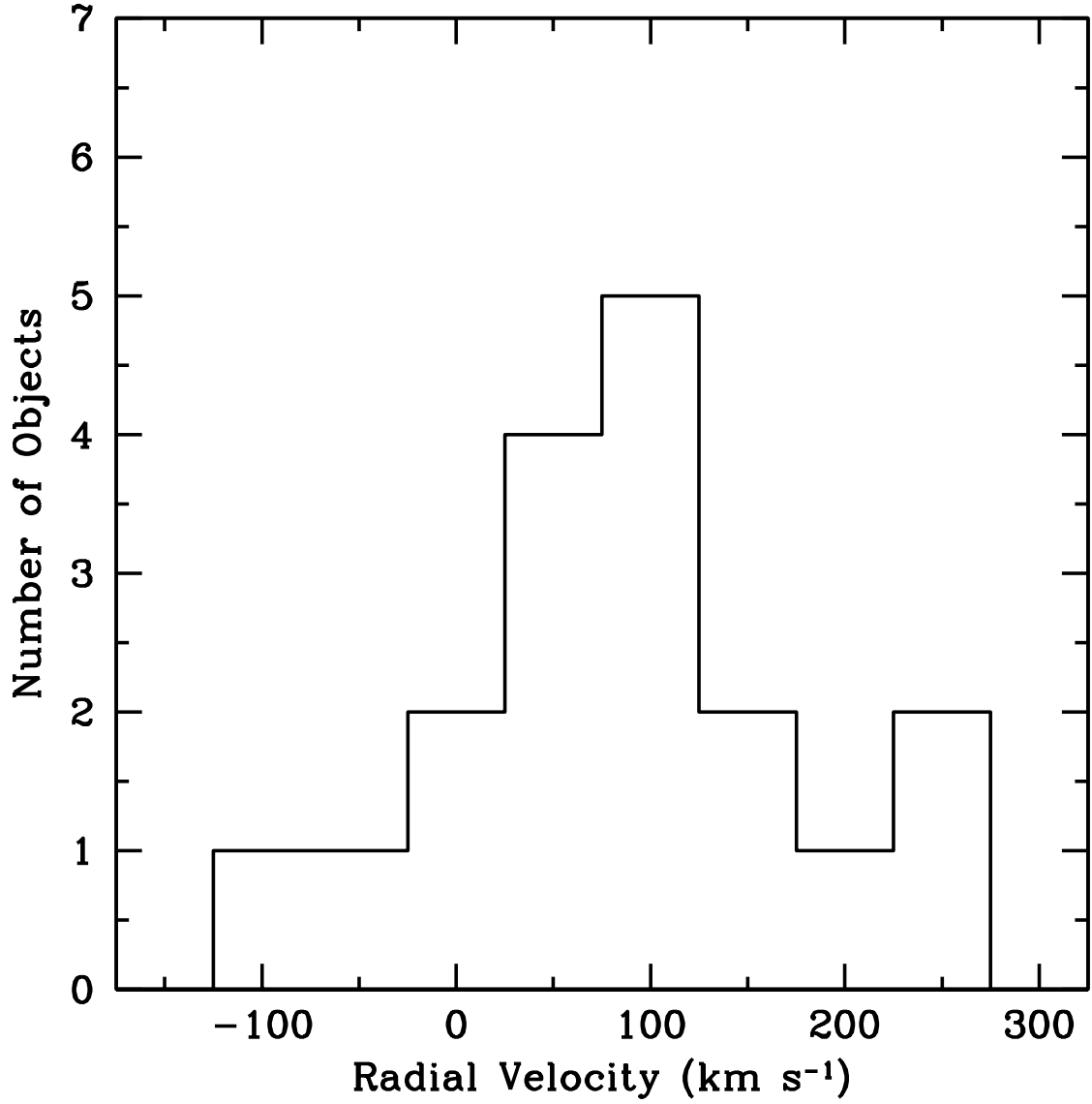}
\caption{Velocity histogram of all NGC 300 GC candidates, including those confirmed by Olsen et al.}
\end{figure}

\begin{figure}
\plotone{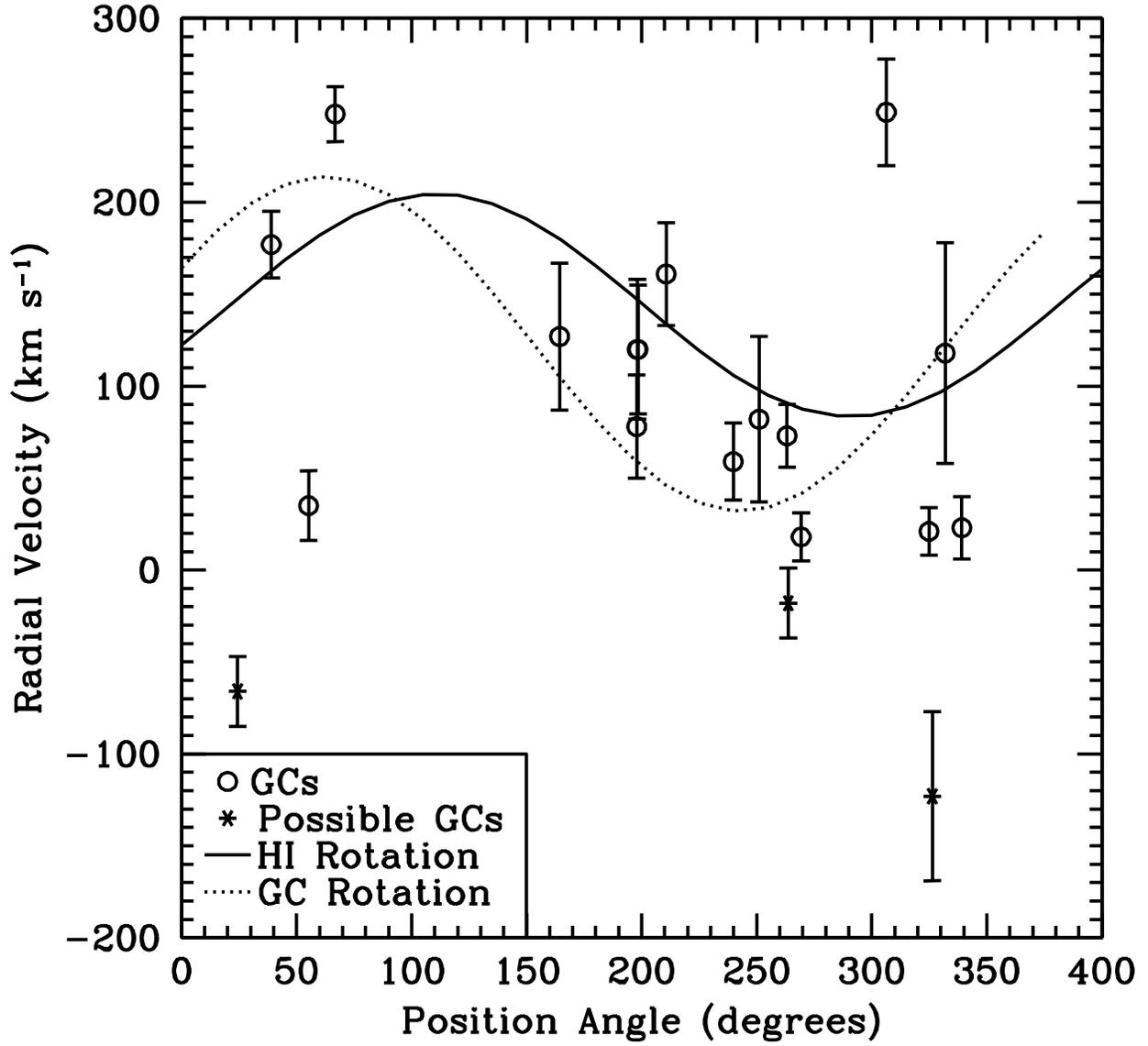}
\caption{Velocity vs. position angle for the 14 highly probable GCs (open circles) and 3 possible GCs (stars), shown with a representation of the H{\small{I}} rotation curve (solid line) and a rotation curve calculated for the 14 highly probable GCs (dotted line).}
\end{figure}

\begin{figure}
\plottwo{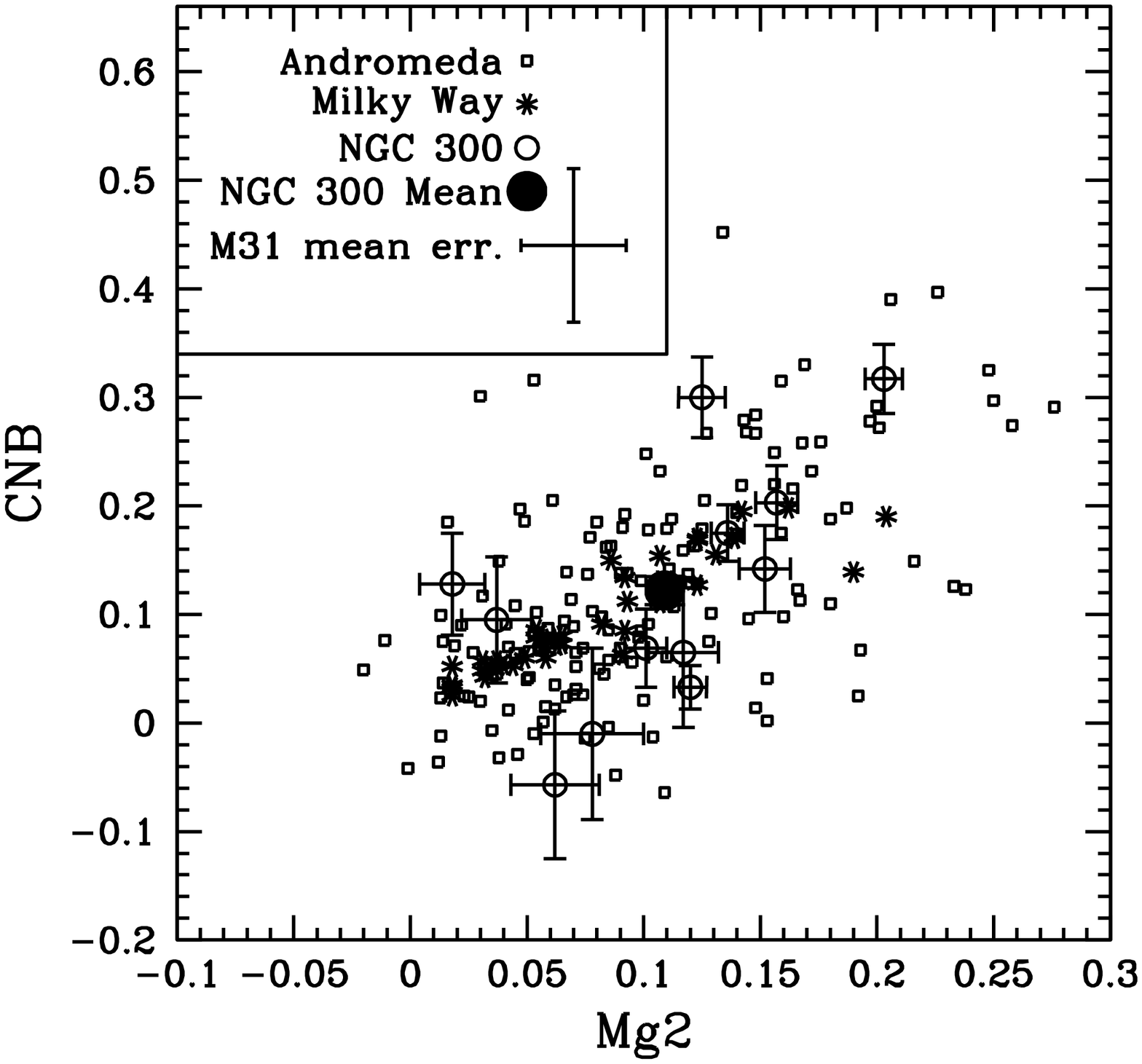}{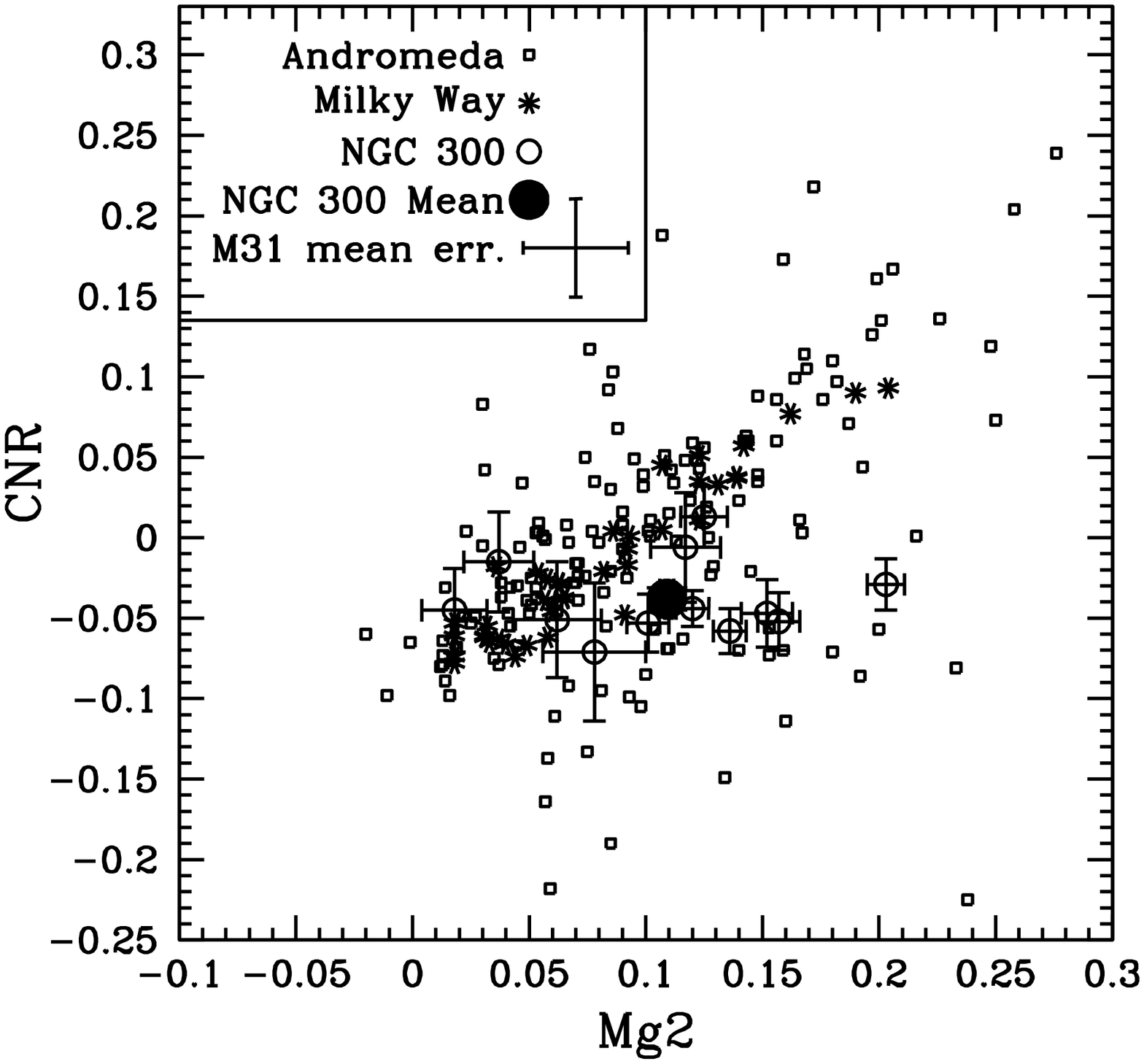}
\caption{CN indices (blue on left, red on right) vs. Mg2 index.  Open squares represent Andromeda clusters, asterisks represent Milky Way clusters, and open circles represent Sculptor Group clusters. A large filled circle marks the mean of the NGC 300 clusters.  The average error bar for M31 is shown in the legend.  The average Milky Way uncertainty was smaller than the size of the symbols and therefore omitted.}
\end{figure}

\begin{figure}
\plotone{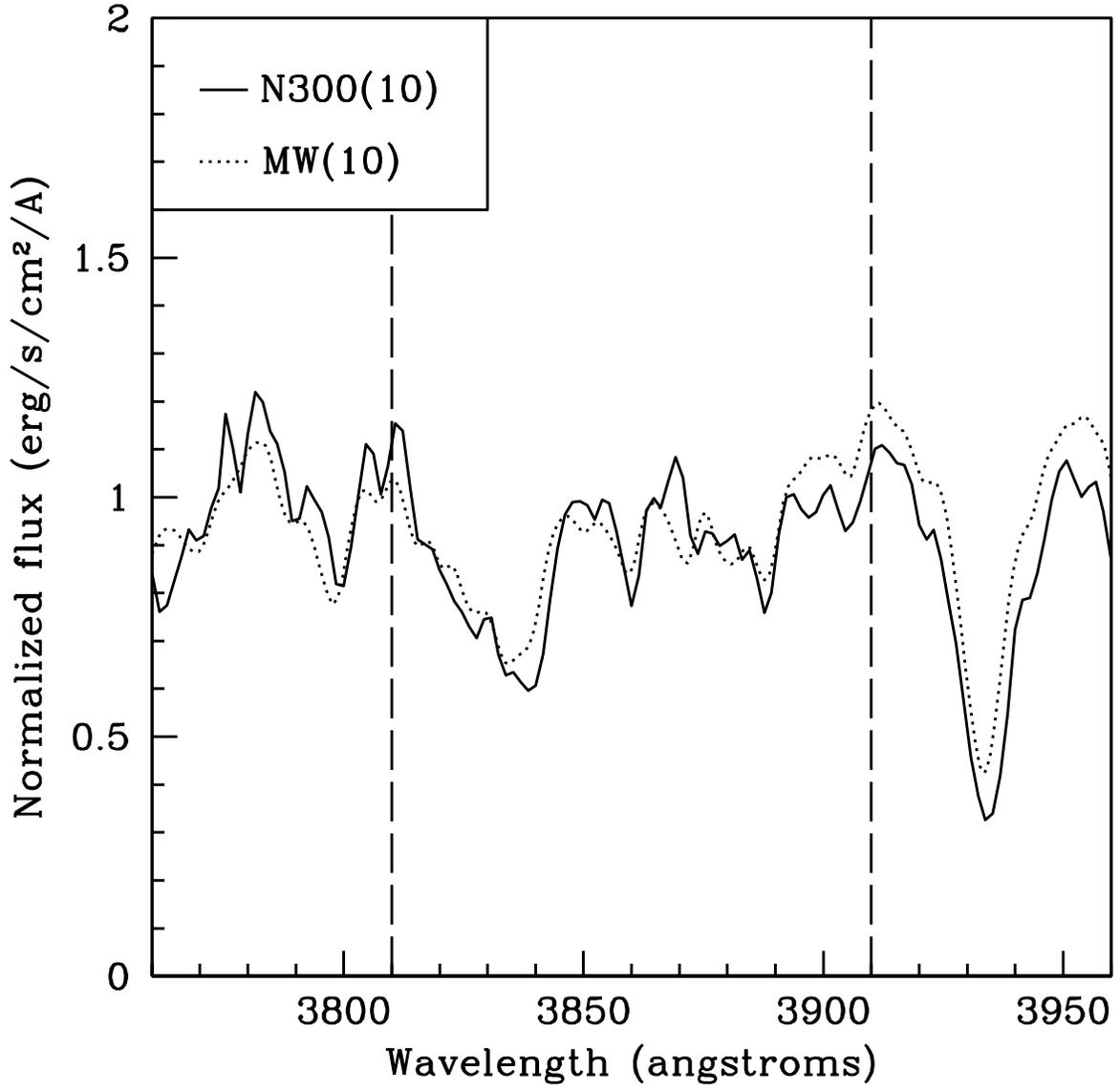}
\caption{CNB feature for averages of 10 NGC 300 GCs and 10 MW GCs matched for Fe52 index strength.}
\end{figure}

\begin{figure}
\plotone{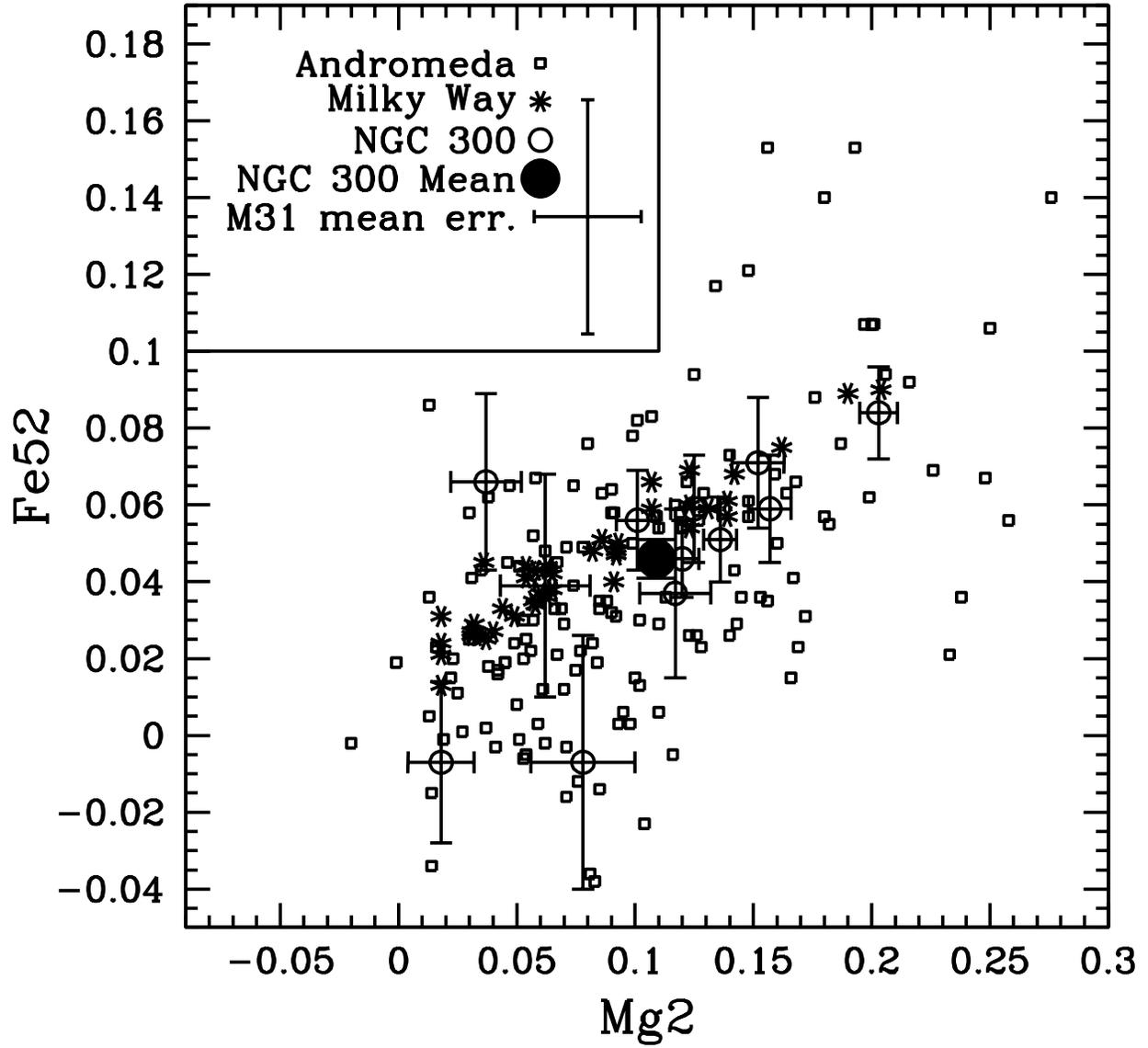}
\caption{Fe-5270 index vs. Mg2 index.  Symbols are the same as in Figure 4.}
\end{figure}

\begin{figure}
\plotone{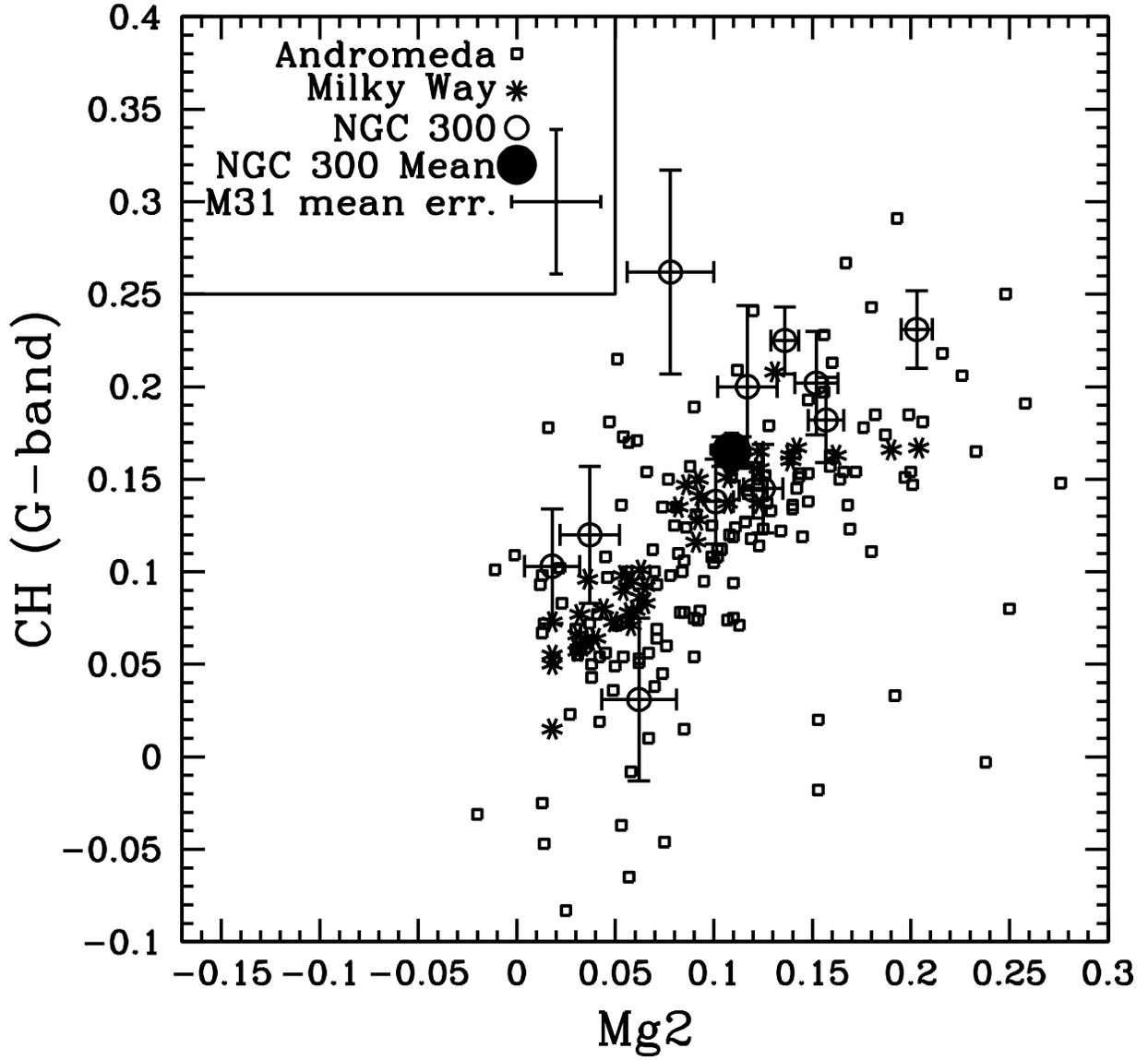}
\caption{G-band (CH) index vs. Mg2 index.  Symbols are the same as in Figure 4.}
\end{figure}

\begin{figure}
\plotone{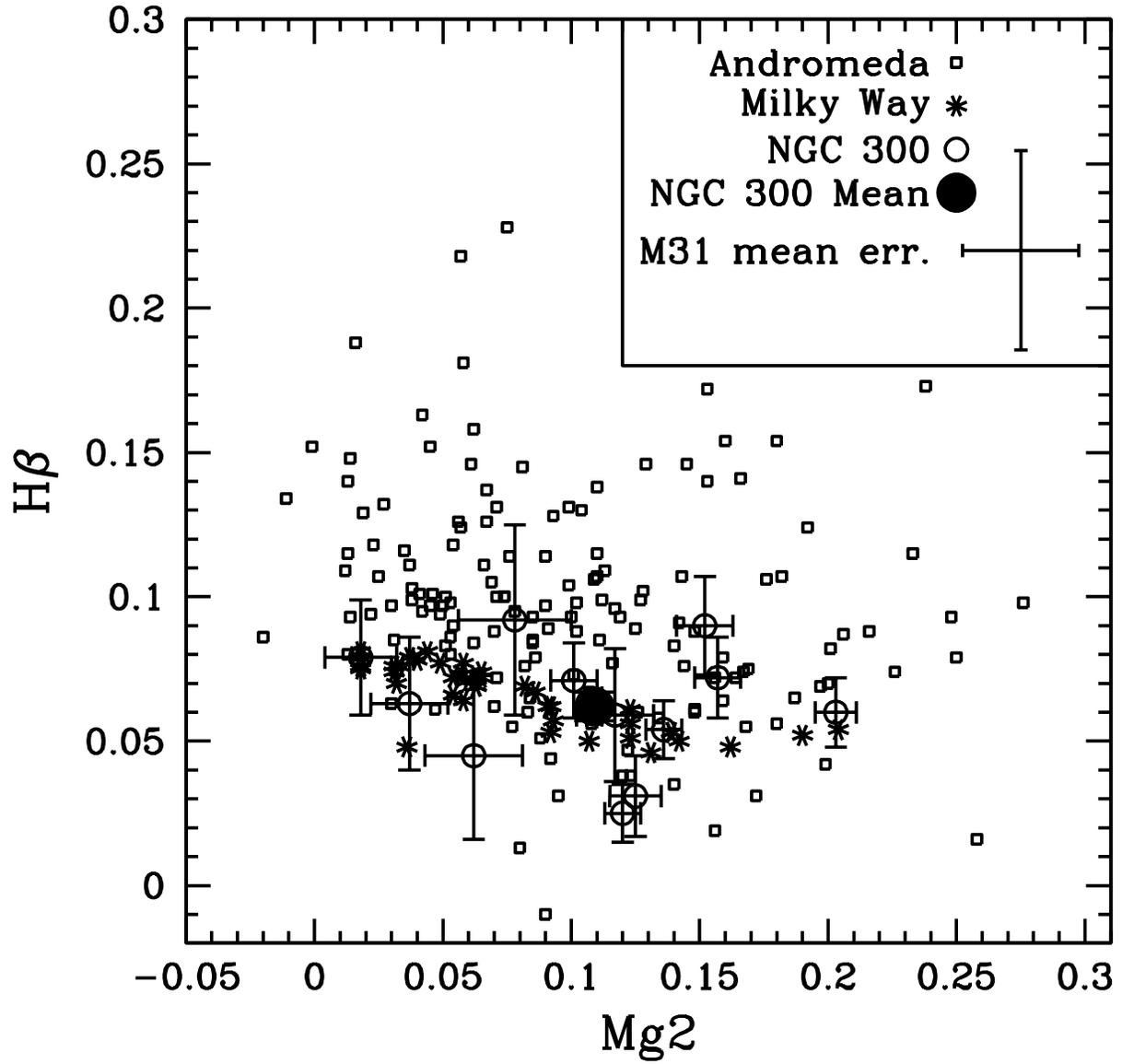}
\caption{H$\beta$ index vs. Mg2 index.  Symbols are the same as in Figure 4.}
\end{figure}

\begin{figure}
\plottwo{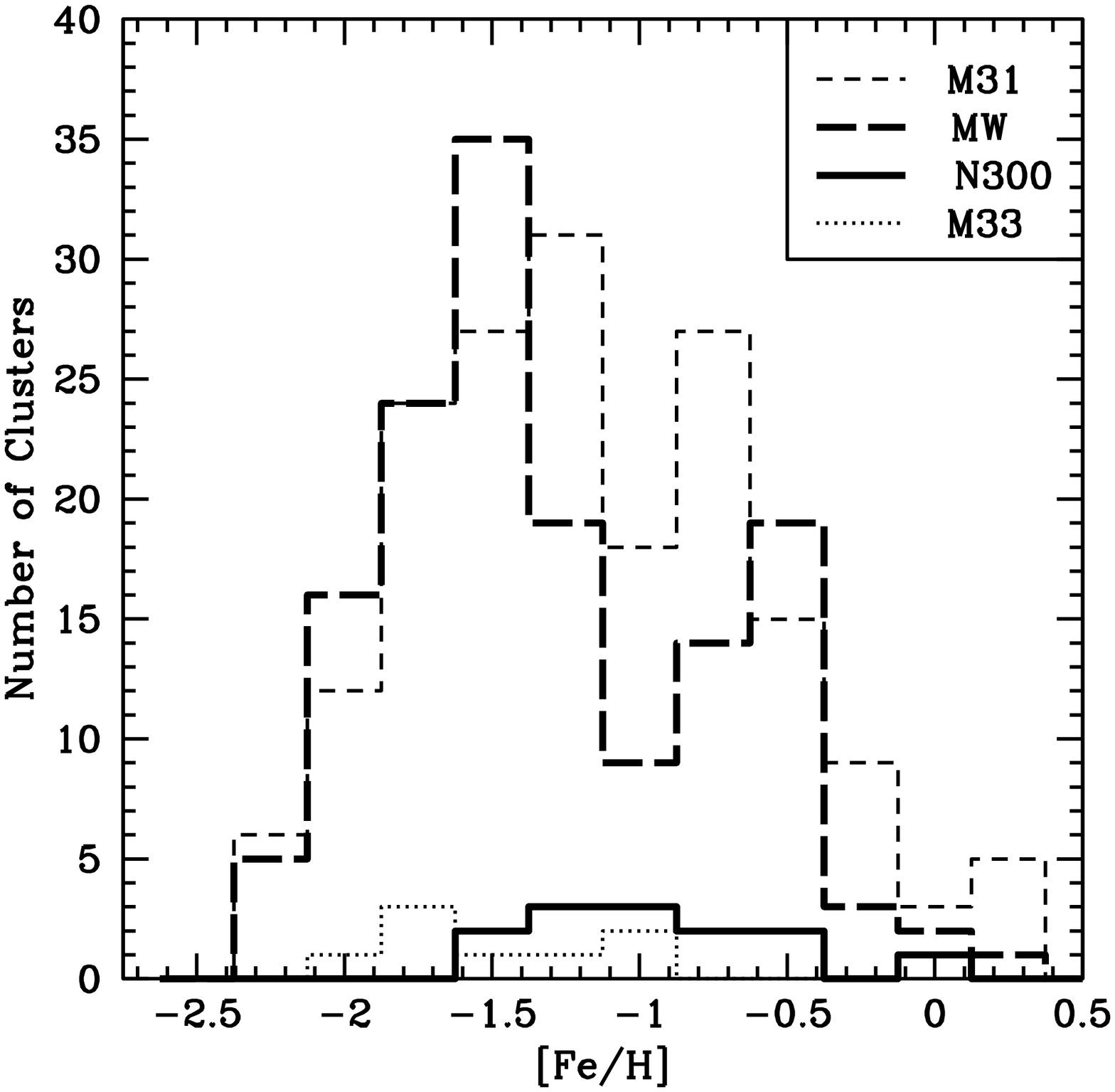}{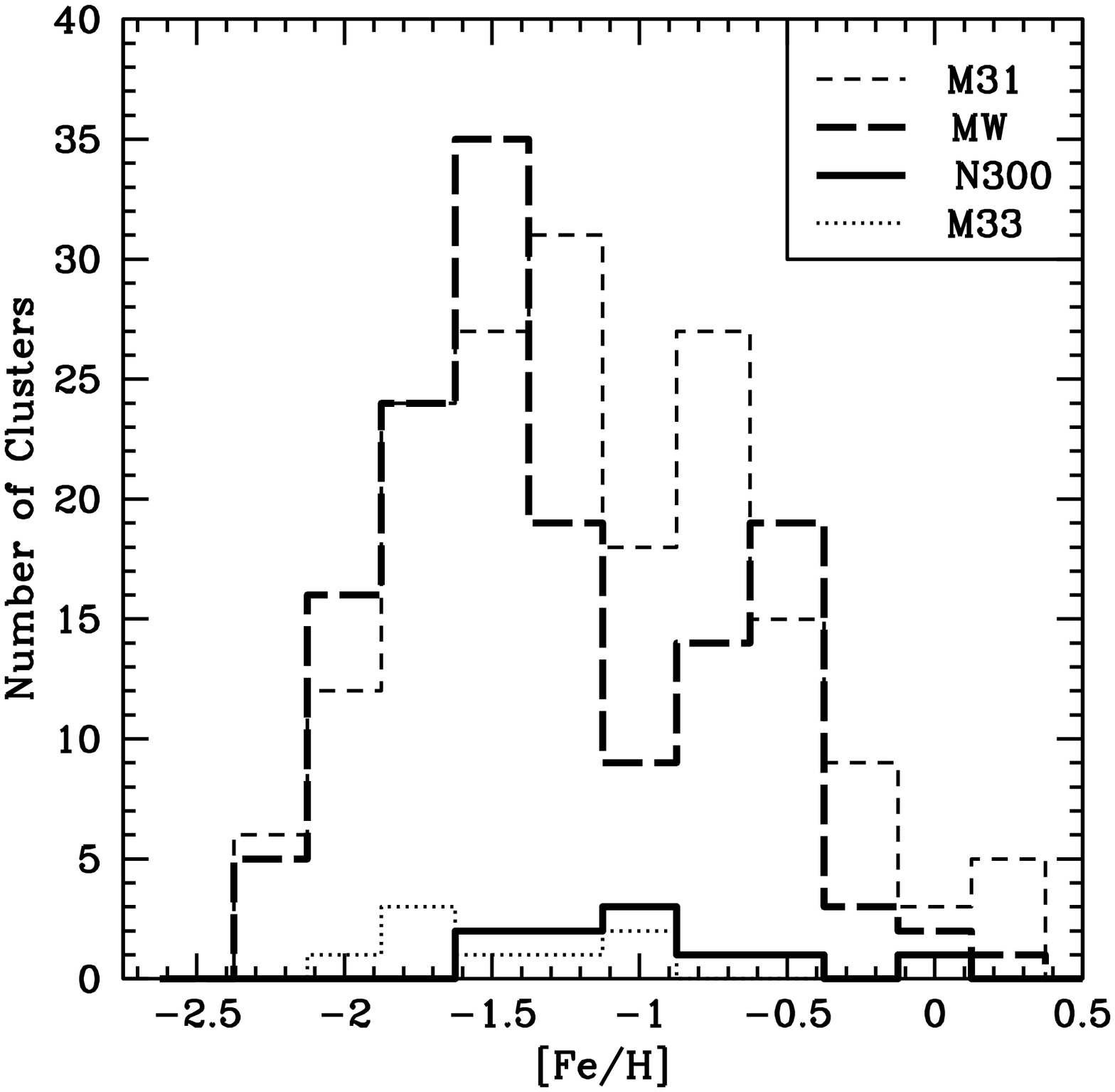}
\caption{NGC 300 metallicity distributions, including (left) and not including (right) candidates flagged as possible stars.  M31, M33, and Milky Way metallicity distributions are shown for comparison.}
\end{figure}

\begin{figure}
\epsscale{1.0}
\plotone{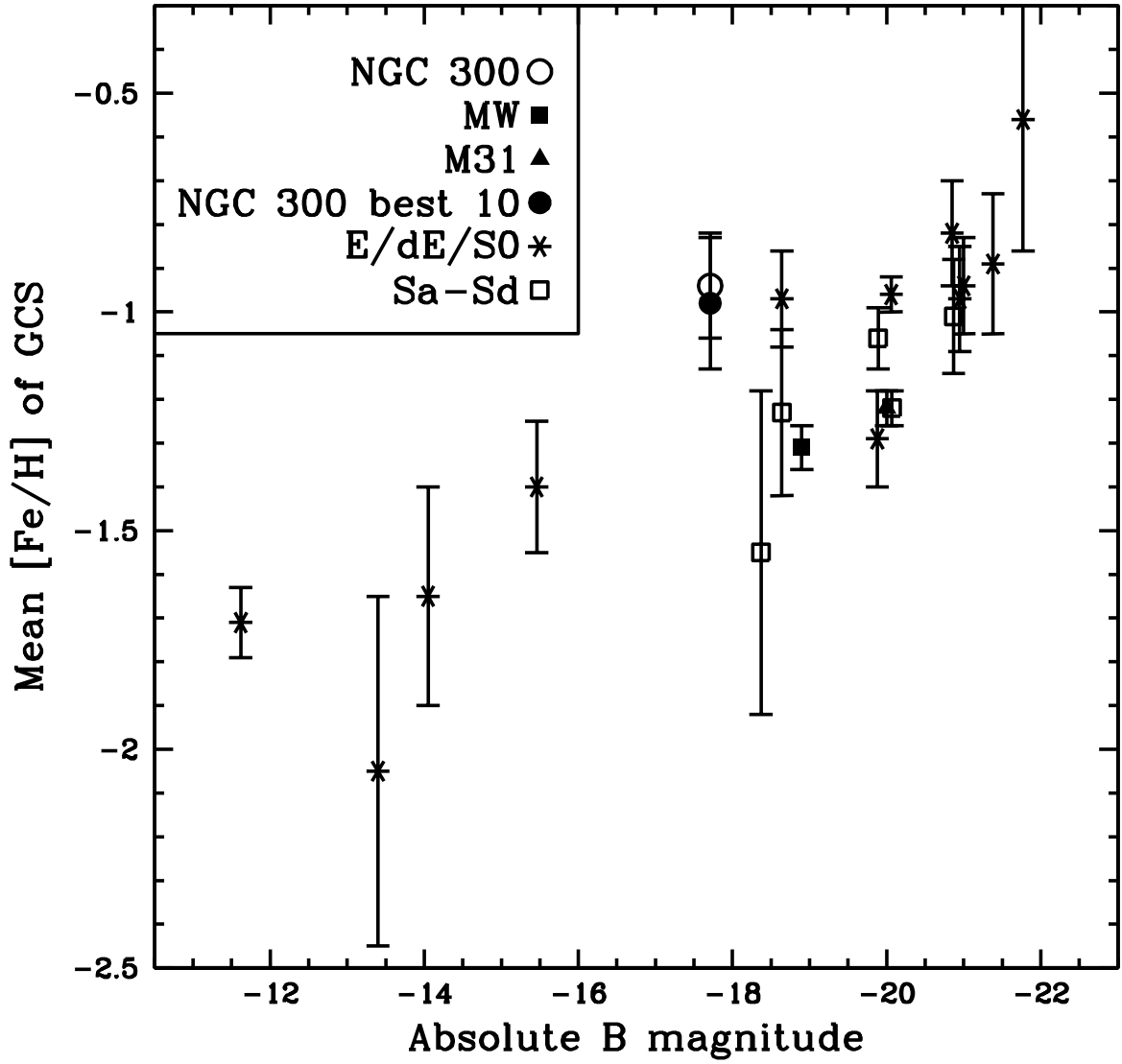}
\caption{Average GCS metallicity vs. absolute B magnitude, showing NGC 300's position including and excluding the low-velocity ``GC or star'' objects.}
\end{figure}

\begin{figure}
\plotone{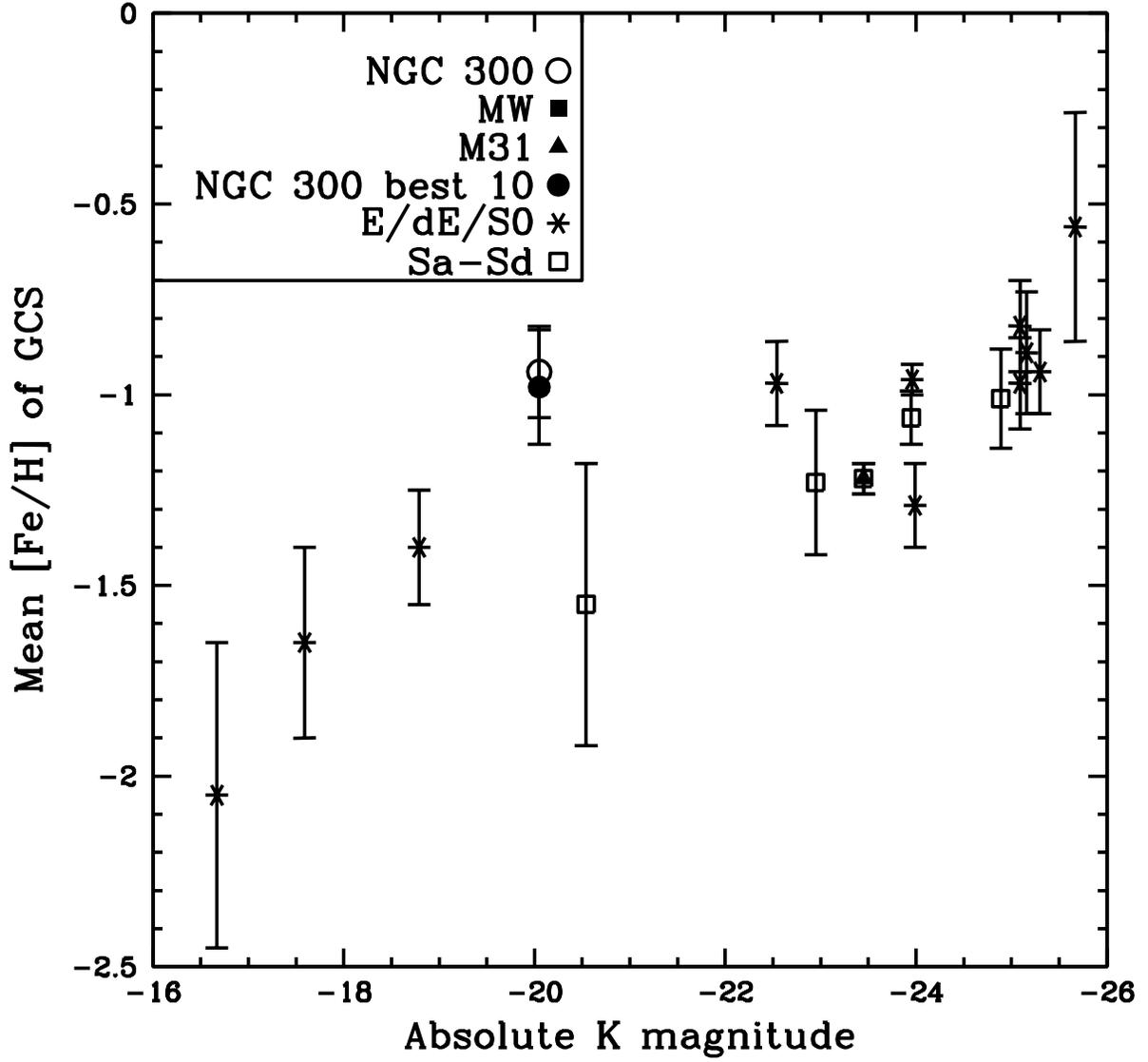}
\caption{Average GCS metallicity vs. absolute K magnitude.  Symbols are the same as in figure 8.}
\end{figure}

\clearpage

\begin{deluxetable}{ccccccccc}


\tabletypesize{\scriptsize}


\tablecaption{Sculptor Velocities and Object Types from Spectra}

\tablehead{\colhead{Name} & \colhead{RA} & \colhead{DEC} & \colhead{K02 class} & \colhead{Velocity} & \colhead{Vel. err.} & R-value & \colhead{Template.} & \colhead{Object Type} \\ 
\colhead{} & \colhead{hh:mm:ss} & \colhead{dd:mm:ss} & \colhead{} & \colhead{(km s$^{-1}$)} & \colhead{(km s$^{-1}$)} & \colhead{} & \colhead{} & \colhead{} } 

\startdata
NGC 253b & 00:46:50.70 & -25:26:38.90 & -- & 209 & 43 & 7.87 & m31\_a\_temp & gc \\
NGC 253ad & 00:47:31.91 & -25:26:42.53 & -- & 65 & 40 & 7.42 & fn4486btemp & gc \\
NGC 300cz & 00:53:26.93 & -37:41:55.54 & -- & 81 & 19 & 16.6 & fallstars & star \\
NGC 300cy & 00:53:27.89 & -37:55:21.48 & -- & 85377 & 24 & 6.73 & hemtemp0.0 & (em.?)gal \\
NGC 300dt & 00:53:40.30 & -37:49:34.38 & -- & 127 & 40 & 8.68 & habtemp90 & gc \\
NGC 300db & 00:53:55.35 & -37:23:42.98 & -- & -123 & 46 & 5.95 & m31\_f\_temp & gc or star \\
NGC 300ds & 00:53:56.12 & -37:34:49.40 & -- & 169 & 21 & 13.27 & fabtemp97 & star \\
NGC 300dd & 00:53:56.47 & -37:58:46.59 & -- & 11483 & 29 & 4.26 & hemtemp0.0 & gal? \\
NGC 300cm & 00:54:05.51 & -37:42:06.11 & -- & -18 & 19 & 14.24 & m31\_k\_temp & gc or star \\
NGC 300-01 & 00:54:22.07 & -37:43:22.1 & 1 & 49374 & 34 & 8.78 & habtemp90 & gal* \\
NGC 300ck & 00:54:21.84 & -37:32:07.82 & -- & 21 & 13 & 19.27 & fallstars & gc \\
NGC 300cj & 00:54:24.13 & -37:25:46.17 & -- & 23 & 17 & 15.63 & m31\_k\_temp & gc \\
NGC 300-08 & 00:54:57.41 & -37:33:55.30 & 2 & 16797 & 22 & 14.43 & habtemp90 & gal* \\
NGC 300-02 & 00:54:35.80 & -37:43:05.60 & 2 & 59 & 21 & 15.11 & habtemp90 & gc \\
NGC 300-15 & 00:55:21.81 & -37:45:40.10 & 2 & 106 & 22 & 20.12 & m31\_a\_temp & star \\
NGC 300-03 & 00:54:40.02 & -37:49:18.71 & 3 & 78 & 28 & 10.04 & m31\_f\_temp & gc \\
NGC 300-04 & 00:54:45.62 & -37:43:41.00 & 2 & 161 & 28 & 9.14 & m31\_f\_temp & gc \\
NGC 300-14 & 00:55:17.67 & -37:36:57.90 & 2 & 16995 & 36 & 8.68 & fallstars & gal \\
NGC 300-05 & 00:54:49.35 & -37:43:30.50 & 2 & 120 & 35 & 8.08 & m31\_f\_temp & gc* \\
NGC 300-09 & 00:54:57.57 & -37:36:37.90 & 2 & 214 & 14 & 18.08 & fglotemp & star* \\
NGC 300-11 & 00:55:01.44 & -37:37:21.80 & 1 & 38350 & 17 & 8.83 & hemtemp0.0 & em.gal* \\
NGC 300cl & 00:54:53.85 & -37:23:57.44 & -- & 21 & 16 & 17.4 & fabtemp97 & star \\
NGC 300dc & 00:54:54.66 & -37:58:34.98 & -- & 68066 & 25 & 5.13 & hemtemp0.0 & em.gal \\
NGC 300df & 00:54:57.02 & -37:58:52.21 & -- & -16 & 48 & 5.76 & m31\_k\_temp & star \\
NGC 300-10 & 00:54:59.03 & -37:36:24.40 & 3 & 173 & 23 & 12.59 & m31\_f\_temp & star* \\
NGC 300ax & 00:55:01.49 & -37:47:36.40 & -- & -27 & 32 & 8.43 & m31\_k\_temp & star \\
NGC 300-07 & 00:54:53.22 & -37:43:11.1 & 1 & 43722 & 27 & 6.41 & hemtemp0.0 & em.gal* \\
NGC 300-12 & 00:55:11.03 & -37:36:46.10 & 1 & 177 & 18 & 15.19 & m31\_f\_temp & gc \\
NGC 300co & 00:55:19.49 & -37:29:38.73 & -- & -66 & 19 & 14.57 & m31\_k\_temp & gc or star \\
NGC 300-06 & 00:54:51.67 & -37:39:38.00 & 3 & 156 & 14 & 8.97 & hemtemp0.0 & HII* \\
NGC 300-17 & 00:55:38.79 & -37:46:30.70 & 3 & 16868 & 21 & 13.02 & fallstars & gal \\
NGC 300-16 & 00:55:27.56 & -37:46:09.40 & 3 & 16798 & 29 & 8.03 & m31\_k\_temp & gal \\
NGC 300-13 & 00:55:11.17 & -37:33:13.90 & 3 & 38182 & 20 & 6.19 & hemtemp0.0 & em.gal \\
NGC 300cr & 00:55:52.72 & -37:59:04.14 & -- & -23 & 15 & 17.05 & fallstars & star \\
NGC 300bd & 00:55:56.79 & -37:52:07.28 & -- & 68818 & 20 & 6.69 & hemtemp0.0 & gal \\
NGC 300cq & 00:56:07.09 & -37:57:02.64 & -- & -79 & 14 & 17.25 & fallstars & star \\
NGC 300cs & 00:56:07.62 & -37:57:23.07 & -- & 31 & 16 & 15.85 & fallstars & star \\
NGC 300da & 00:56:09.49 & -37:31:21.25 & -- & 49619 & 21 & 7.51 & hemtemp0.0 & em.gal \\
NGC 300cv & 00:56:13.97 & -37:34:16.27 & -- & 97 & 17 & 12.33 & fabtemp97 & star \\
NGC 300ct & 00:56:18.24 & -37:28:48.08 & -- & 4 & 15 & 17.59 & fabtemp97 & star \\
NGC 300cp & 00:56:19.03 & -37:29:17.23 & -- & 35 & 19 & 13.88 & m31\_k\_temp & gc \\
NGC 300cx & 00:56:20.30 & -37:59:11.68 & -- & 68618 & 53 & 5.97 & m31\_f\_temp & gal \\
NGC 300cn & 00:56:23.52 & -37:27:37.57 & -- & 49828 & 17 & 8.89 & hemtemp0.0 & em.gal \\
NGC 300cu & 00:56:24.62 & -37:52:56.29 & -- & 18 & 20 & 12.33 & fallstars & star \\
\enddata

\tablecomments{Objects with letter designations (e.g. NGC 300cx) are from \citet{ols04}.  Objects with number designations (e.g. NGC 300-05) are from \citet{kim02}.  Objects with asterisks next to the object type have been visually inspected in archival HST images.}





\end{deluxetable}

\begin{deluxetable}{ccccccc}


\tabletypesize{\scriptsize}


\tablecaption{Radial Velocities of Previously Observed GCs}


\tablehead{\colhead{ID} & \colhead{RA (J2000)} & \colhead{Dec (J2000)} & \colhead{RV} & \colhead{$\sigma_{RV}$} & \colhead{Object type} & \colhead{Vel. source} \\ 
\colhead{} & \colhead{(hours)} & \colhead{(minutes)} & \colhead{(km s$^{-1}$)} & \colhead{(km s$^{-1}$)} & \colhead{} & \colhead{} } 

\startdata
NGC 300ac & 00:54:02.48 & -37:44:31.64 & 82 & 45 & gc & Olsen et al. 2004 \\
NGC 300ag & 00:54:23.22 & -37:59:17.67 & 120 & 38 & gc & Olsen et al. 2004 \\
NGC 300am & 00:54:07.54 & -37:42:09.23 & 73 & 17 & gc & Olsen (unpublished) \\
NGC 300ba & 00:54:05.49 & -37:23:11.24 & 118 & 60 & gc & Olsen et al. 2004 \\
NGC 300m & 00:54:07.04 & -37:41:10.35 & 18 & 13 & gc & Olsen (unpublished) \\
NGC 300r & 00:56:23.13 & -37:33:26.98 & 248 & 15 & gc & Olsen et al. 2004 \\
NGC 300s & 00:53:55.66 & -37:32:37.38 & 249 & 29 & gc & Olsen et al. 2004 \\

\enddata




\end{deluxetable}

\begin{deluxetable}{cccccccccccccc}

\tabletypesize{\scriptsize}


\tablecaption{Brodie \& Huchra Calibrated Spectral Indices and H$\delta$}

\tablehead{\colhead{Name} & \colhead{CNR} & \colhead{CH/G} & \colhead{H$\beta$} & \colhead{MgH} & \colhead{Mg2} & \colhead{Mgb} & \colhead{Fe52} & \colhead{NaI} & \colhead{CNB} & \colhead{H\&K} & \colhead{MgG} & \colhead{$\delta$} &\colhead{H$\delta$A}\\ 
\colhead{} & \colhead{mag} & \colhead{mag} & \colhead{mag} & \colhead{mag} & \colhead{mag} & \colhead{mag} & \colhead{mag} & \colhead{mag} & \colhead{mag} & \colhead{mag} & \colhead{mag} & \colhead{mag} &\colhead{mag}}
\startdata
NGC 253ad  & 0.017 & 0.262 & 0.039 & 0.048 & 0.177 & 0.152 & 0.027 & 0.046 & 0.294 & 0.312 & -0.086 & 0.477 & -0.83 \\
$\sigma$  & 0.060 & 0.071 & 0.037 & 0.022 & 0.025 & 0.043 & 0.035 & 0.041 & 0.120 & 0.121 & 0.069 & 0.031 & 0.066\\
NGC 253b  & -0.132 & 0.037 & 0.058 & 0.009 & 0.017 & 0.005 & -0.005 & 0.049 & 0.216 & 0.234 & -0.088 & 0.252 & 0.101 \\
$\sigma$  & 0.042 & 0.053 & 0.033 & 0.020 & 0.022 & 0.037 & 0.033 & 0.042 & 0.071 & 0.067 & 0.050 & 0.019 & 0.045 \\
NGC 300-02  & -0.044 & 0.144 & 0.025 & 0.001 & 0.120 & 0.110 & 0.046 & 0.039 & 0.033 & 0.216 & 0.001 & 0.244 & 0.001 \\
$\sigma$  & 0.011 & 0.015 & 0.010 & 0.006 & 0.007 & 0.012 & 0.010 & 0.013 & 0.020 & 0.019 & 0.015 & 0.005 & 0.012\\
NGC 300-03  & -0.045 & 0.103 & 0.079 & -0.008 & 0.018 & 0.045 & -0.007 & 0.035 & 0.128 & 0.312 & -0.188 & 0.298 & 0.048 \\
$\sigma$   & 0.026 & 0.031 & 0.020 & 0.013 & 0.014 & 0.023 & 0.021 & 0.029 & 0.047 & 0.044 & 0.035 & 0.012 & 0.028 \\
NGC 300-04  & -0.015 & 0.120 & 0.063 & 0.010 & 0.037 & 0.064 & 0.066 & 0.008 & 0.095 & 0.328 & -0.165 & 0.354 & 0.013 \\
$\sigma$   & 0.031 & 0.037 & 0.023 & 0.014 & 0.015 & 0.026 & 0.023 & 0.030 & 0.058 & 0.057 & 0.040 & 0.015 & 0.034 \\
NGC 300-05  & -0.051 & 0.031 & 0.045 & 0.007 & 0.062 & 0.086 & 0.039 & -0.047 & -0.057 & 0.184 & -0.134 & 0.243 & 0.090 \\
$\sigma$   & 0.036 & 0.044 & 0.029 & 0.017 & 0.019 & 0.033 & 0.029 & 0.043 & 0.068 & 0.062 & 0.051 & 0.016 & 0.038 \\
NGC 300-12  & -0.053 & 0.138 & 0.071 & 0.030 & 0.101 & 0.088 & 0.056 & 0.026 & 0.069 & 0.362 & -0.245 & 0.446 & 0.024 \\
$\sigma$   & 0.018 & 0.023 & 0.013 & 0.008 & 0.009 & 0.015 & 0.013 & 0.016 & 0.036 & 0.034 & 0.026 & 0.009 & 0.020 \\
NGC 300cj  & -0.058 & 0.225 & 0.054 & 0.012 & 0.136 & 0.136 & 0.051 & 0.049 & 0.175 & 0.399 & -0.187 & 0.536 & -0.043 \\
$\sigma$  & 0.014 & 0.018 & 0.010 & 0.006 & 0.007 & 0.012 & 0.011 & 0.012 & 0.026 & 0.025 & 0.018 & 0.007 & 0.014 \\
NGC 300ck  & -0.029 & 0.231 & 0.060 & 0.030 & 0.203 & 0.180 & 0.084 & 0.070 & 0.317 & 0.378 & -0.200 & 0.664 & -0.62 \\
$\sigma$  & 0.016 & 0.021 & 0.012 & 0.007 & 0.008 & 0.014 & 0.012 & 0.013 & 0.032 & 0.029 & 0.023 & 0.008 & 0.017 \\
NGC 300cm  & -0.052 & 0.182 & 0.072 & 0.022 & 0.157 & 0.160 & 0.059 & 0.036 & 0.203 & 0.350 & -0.108 & 0.483 & -0.023 \\
$\sigma$  & 0.018 & 0.023 & 0.014 & 0.008 & 0.009 & 0.016 & 0.014 & 0.016 & 0.034 & 0.033 & 0.023 & 0.009 & 0.019\\
NGC 300co  & 0.013 & 0.145 & 0.031 & 0.013 & 0.125 & 0.121 & 0.059 & 0.060 & 0.300 & 0.434 & -0.181 & 0.561 & -0.049 \\
$\sigma$  & 0.019 & 0.024 & 0.014 & 0.009 & 0.010 & 0.017 & 0.014 & 0.016 & 0.037 & 0.035 & 0.025 & 0.009 & 0.020 \\
NGC 300cp  & -0.047 & 0.202 & 0.090 & -0.001 & 0.152 & 0.166 & 0.071 & 0.027 & 0.142 & 0.375 & -0.082 & 0.460 & 0.005 \\
$\sigma$  & 0.021 & 0.028 & 0.017 & 0.010 & 0.011 & 0.019 & 0.017 & 0.020 & 0.040 & 0.039 & 0.028 & 0.010 & 0.022 \\
NGC 300db  & -0.071 & 0.262 & 0.092 & 0.002 & 0.078 & 0.114 & -0.007 & -0.026 & -0.010 & 0.268 & -0.022 & 0.226 & 0.002 \\
$\sigma$  & 0.043 & 0.055 & 0.033 & 0.020 & 0.022 & 0.037 & 0.033 & 0.042 & 0.079 & 0.076 & 0.056 & 0.019 & 0.046 \\
NGC 300dt  & -0.006 & 0.200 & 0.059 & 0.008 & 0.117 & 0.177 & 0.037 & 0.087 & 0.065 & 0.268 & -0.024 & 0.269 & -0.052 \\
$\sigma$  & 0.034 & 0.044 & 0.023 & 0.013 & 0.015 & 0.026 & 0.022 & 0.028 & 0.069 & 0.065 & 0.046 & 0.017 & 0.037 \\
$<$NGC 300$>$ & -0.038 & 0.165 & 0.062 & 0.011 & 0.109 & 0.120 & 0.046 & 0.030 & 0.122 & 0.323 & -0.128 & 0.399 & -0.004 \\
$\sigma$  & 0.007 & 0.009 & 0.005 & 0.003 & 0.004 & 0.006 & 0.005 & 0.007 & 0.013 & 0.012 & 0.009 & 0.003 & 0.026 \\
\enddata


\tablecomments{CNR $\sim$ Lick/IDS CN1; MgH $\sim$ Lick/IDS Index Mg1; Na1 $\sim$Lick/IDS index NaD.  All other indices correspond to the Lick/IDS indices of the same name.  CNB, H\&K, MgG, and $\delta$ have no Lick/IDS equivalents..}


\end{deluxetable}
\begin{deluxetable}{cccccccccccccc}

\tabletypesize{\scriptsize}


\tablecaption{Other Spectral Indices}

\tablehead{\colhead{Name} & \colhead{CN2} & \colhead{Ca42} & \colhead{Fe43} & \colhead{Ca44} & \colhead{Fe45} & \colhead{C2} & \colhead{Fe50} & \colhead{Fe53} & \colhead{Fe54} & \colhead{Fe5709} & \colhead{Fe5782} & \colhead{TiO1} & \colhead{Ti02} \\ 
\colhead{} & \colhead{mag} & \colhead{mag} & \colhead{mag} & \colhead{mag} & \colhead{mag} & \colhead{mag} & \colhead{mag} & \colhead{mag} & \colhead{mag} & \colhead{mag} & \colhead{mag} & \colhead{mag} & \colhead{mag}}
\startdata
NGC 253ad  & 0.041 & 0.102 & 0.088 & 0.048 & 0.056 & 0.060 & 0.052 & 0.036 & 0.036 & 0.058 & -0.014 & 0.008 & 0.028  \\
$\sigma$ & 0.076 & 0.106 & 0.066 & 0.067 & 0.055 & 0.042 & 0.037 & 0.044 & 0.045 & 0.038 & 0.045 & 0.024 & 0.023  \\
NGC 253b  & -0.104 & -0.048 & 0.001 & 0.049 & 0.075 & 0.067 & 0.001 & -0.029 & -0.029 & 0.001 & -0.021 & -0.010 & 0.008  \\
$\sigma$ & 0.055 & 0.076 & 0.052 & 0.053 & 0.046 & 0.036 & 0.033 & 0.042 & 0.042 & 0.037 & 0.043 & 0.024 & 0.023  \\
NGC 300-02  & -0.030 & 0.066 & 0.067 & 0.032 & 0.046 & 0.013 & 0.032 & 0.047 & 0.047 & 0.018 & 0.022 & 0.009 & 0.012  \\
$\sigma$ & 0.015 & 0.020 & 0.014 & 0.015 & 0.013 & 0.011 & 0.010 & 0.013 & 0.014 & 0.012 & 0.015 & 0.008 & 0.008  \\
NGC 300-03  & -0.021 & 0.072 & -0.020 & 0.023 & 0.081 & 0.024 & 0.046 & -0.008 & -0.008 & 0.019 & -0.019 & -0.032 & 0.007  \\
$\sigma$ & 0.034 & 0.045 & 0.032 & 0.032 & 0.028 & 0.022 & 0.021 & 0.027 & 0.027 & 0.024 & 0.029 & 0.016 & 0.015  \\
NGC 300-04  & -0.026 & 0.072 & 0.060 & 0.028 & 0.069 & 0.044 & 0.065 & 0.052 & 0.052 & 0.038 & 0.024 & 0.007 & 0.011  \\
$\sigma$ & 0.042 & 0.055 & 0.038 & 0.038 & 0.032 & 0.025 & 0.023 & 0.028 & 0.029 & 0.027 & 0.032 & 0.018 & 0.017  \\
NGC 300-05  & -0.052 & 0.081 & -0.004 & -0.045 & 0.003 & 0.012 & -0.017 & 0.031 & 0.031 & -0.025 & -0.047 & 0.005 & 0.016  \\
$\sigma$ & 0.049 & 0.062 & 0.045 & 0.047 & 0.040 & 0.031 & 0.030 & 0.036 & 0.037 & 0.036 & 0.043 & 0.024 & 0.023  \\
NGC 300-12  & -0.023 & 0.102 & 0.064 & 0.030 & 0.069 & 0.027 & 0.036 & 0.040 & 0.040 & 0.023 & 0.022 & 0.042 & 0.064  \\
$\sigma$ & 0.024 & 0.032 & 0.021 & 0.023 & 0.018 & 0.015 & 0.014 & 0.017 & 0.017 & 0.015 & 0.018 & 0.010 & 0.010  \\
NGC 300cj  & -0.036 & 0.069 & 0.096 & 0.050 & 0.067 & 0.039 & 0.050 & 0.041 & 0.041 & 0.027 & 0.013 & -0.009 & 0.009  \\
$\sigma$ & 0.018 & 0.024 & 0.017 & 0.017 & 0.014 & 0.012 & 0.011 & 0.013 & 0.014 & 0.012 & 0.014 & 0.008 & 0.008  \\
NGC 300ck  & 0.011 & 0.147 & 0.120 & 0.069 & 0.072 & 0.050 & 0.068 & 0.077 & 0.077 & 0.016 & 0.030 & 0.001 & 0.017  \\
$\sigma$ & 0.020 & 0.028 & 0.018 & 0.019 & 0.016 & 0.013 & 0.012 & 0.015 & 0.016 & 0.013 & 0.016 & 0.009 & 0.009  \\
NGC 300cm  & -0.026 & 0.144 & 0.083 & 0.071 & 0.053 & 0.004 & 0.026 & 0.052 & 0.052 & 0.001 & 0.030 & 0.020 & 0.005  \\
$\sigma$ & 0.023 & 0.032 & 0.021 & 0.022 & 0.019 & 0.015 & 0.014 & 0.017 & 0.018 & 0.016 & 0.019 & 0.010 & 0.010  \\
NGC 300co  & 0.039 & 0.124 & 0.039 & 0.041 & 0.049 & 0.012 & 0.044 & 0.039 & 0.039 & 0.023 & 0.019 & -0.001 & 0.008  \\
$\sigma$ & 0.024 & 0.034 & 0.022 & 0.024 & 0.020 & 0.016 & 0.015 & 0.018 & 0.019 & 0.016 & 0.019 & 0.010 & 0.010  \\
NGC 300cp  & -0.009 & 0.110 & 0.039 & 0.045 & 0.055 & 0.038 & 0.037 & 0.044 & 0.044 & 0.001 & 0.023 & -0.008 & 0.015  \\
$\sigma$ & 0.027 & 0.038 & 0.026 & 0.027 & 0.022 & 0.018 & 0.017 & 0.021 & 0.022 & 0.019 & 0.023 & 0.012 & 0.012  \\
NGC 300db  & -0.043 & 0.091 & -0.023 & -0.004 & -0.061 & 0.016 & 0.022 & 0.043 & 0.043 & -0.025 & -0.080 & -0.058 & -0.021  \\
$\sigma$ & 0.056 & 0.074 & 0.055 & 0.057 & 0.048 & 0.037 & 0.034 & 0.041 & 0.042 & 0.038 & 0.046 & 0.024 & 0.025  \\
NGC 300dt  & 0.000 & 0.131 & 0.134 & 0.088 & 0.063 & -0.006 & 0.037 & 0.009 & 0.009 & 0.025 & -0.048 & -0.021 & 0.002  \\
$\sigma$ & 0.044 & 0.056 & 0.036 & 0.037 & 0.032 & 0.025 & 0.022 & 0.028 & 0.028 & 0.025 & 0.030 & 0.016 & 0.016  \\
$<$NGC 300$>$ & -0.018 & 0.101 & 0.055 & 0.036 & 0.047 & 0.023 & 0.037 & 0.039 & 0.037 & 0.012 & -0.001 & -0.004 & 0.012  \\
$\sigma$ & 0.009 & 0.012 & 0.008 & 0.009 & 0.007 & 0.006 & 0.005 & 0.007 & 0.007 & 0.006 & 0.007 & 0.004 & 0.004  \\
\enddata




\end{deluxetable}

\begin{deluxetable}{ccccccc}


\tablecaption{Linear Fits to $[Fe/H] = a(index)+b$}


\tablehead{\colhead{Index ID} & \colhead{a} & \colhead{b} & \colhead{$R^2$} & \colhead{$R_I$} & \colhead{$\sigma_m$} & \colhead{$\sigma_s$} \\ 
\colhead{} & \colhead{(dex/mag)} & \colhead{(dex)} & \colhead{} & \colhead{} & \colhead{} & \colhead{}} 


\startdata
Ca4455 & 25.85 & -2.00 & 0.899 & 0.0387 & 0.006 & 0.003\\
Fe5270 & 26.50 & -2.31 & 0.888 & 0.0377 & 0.006 & 0.001\\
Mg2 & 9.83 & -1.89 & 0.886 & 0.1017 & 0.017 & 0.006\\
$\delta$ & 4.12 & -2.68 & 0.884 & 0.2428 & 0.041 & 0.095\\
Fe5335 & 30.36 & -2.22 & 0.879 & 0.0329 & 0.006 & 0.005\\
Ca4227 & 18.65 & -2.09 & 0.878 & 0.0536 & 0.010 & 0.014\\
CN2 & 9.45 & -1.27 & 0.852 & 0.1058 & 0.020 & 0.021\\
G4300 & 10.37 & -2.24 & 0.844 & 0.0964 & 0.019 & 0.008\\
\enddata


\end{deluxetable}

\begin{deluxetable}{ccccc}




\tablecaption{Sculptor GC Metallicities}

\tablehead{\colhead{Name} & \colhead{[Fe/H]} & \colhead{$\sigma_{[Fe/H]}$} & \colhead{[Fe/H] Src.} & \colhead{Ols. No.} \\ 
\colhead{} & \colhead{dex} & \colhead{dex} & \colhead{} & \colhead{} } 

\startdata
NGC 55az & -1.76 & 0.43 & Ols & 27 \\
NGC 253q & -1.07 & 0.30 & Ols & 5 \\
NGC 253b & -1.96 & 0.54 & new & 26 \\
NGC 247a & -1.04 & 0.29 & Ols & 64 \\
NGC 253a & -0.88 & 0.25 & Ols & 46 \\
NGC 253ad & -0.52 & 0.56 & new & 57 \\
NGC 300dt & -0.89 & 0.66 & new & 13 \\
NGC 300db & -1.23 & 0.81 & new & 22 \\
NGC 300cm & -0.54 & 0.36 & new & 33 \\
NGC 300ck & -0.07 & 0.25 & new & 42 \\
NGC 300cj & -0.71 & 0.35 & new & 45 \\
NGC 300-02 & -1.03 & 0.29 & new & n/a \\
NGC 300-03 & -1.61 & 0.47 & new & 48 \\
NGC 300-04 & -1.18 & 0.33 & new & n/a \\
NGC 300-05 & -1.57 & 0.52 & new & n/a \\
NGC 300-12 & -0.93 & 0.23 & new & n/a \\
NGC 300co & -0.67 & 0.29 & new & 69 \\
NGC 300cp & -0.59 & 0.29 & new & 116 \\
NGC 300r & -1.25 & 0.35 & Ols & 121 \\
\enddata


\tablecomments{Ols.No. is the object number in \citet{ols04} Table 4.  Src. designates the source of the metallicity value: "Ols" for \citet{ols04} or "new" for the newly reduced spectra.}


\end{deluxetable}

\begin{deluxetable}{ccccccccccc}


\tabletypesize{\scriptsize}


\tablecaption{Mean Galaxy GCS Metallicities and Luminosities}


\tablehead{\colhead{ID} & \colhead{$<[m/H]>$} & \colhead{$\sigma_{<[m/H]>}$} & \colhead{$B_t$} & \colhead{Distance} & \colhead{$M_{Bt}$} & \colhead{$K$} & \colhead{$M_K$} & \colhead{Ref-Metal} & \colhead{Ref-Dist.} & \colhead{Ref-K} \\ 
\colhead{} & \colhead{(dex)} & \colhead{(dex)} & \colhead{(mag)} & \colhead{(Mpc)} & \colhead{(mag)} & \colhead{(mag)} & \colhead{(mag)} & \colhead{} & \colhead{} & \colhead{} } 

\startdata
Fornax & -1.71 & 0.08 & 9.28 & 0.136 & -11.38 & -- & -- & 32 & 5 & -- \\
M87 & -0.89 & 0.16 & 9.56 & 15.631 & -21.38 & 5.81 & -25.16 & 6 & 11 & 17 \\
M31 & -1.22 & 0.04 & 4.36 & 0.77 & -20.07 & 0.98 & -23.45 & 2 & 12 & 17 \\
MW & -1.31 & 0.05 & 5.52 & 0.76 & -18.9 & -- & -- & 16 & -- & -- \\
M81 & -1.19 & 0.14 & 7.89 & 3.6 & -19.89 & 3.83 & -23.95 & 29, 26 & 13 & 17 \\
M33 & -1.55 & 0.37 & 6.27 & 0.847 & -18.37 & 4.1 & -20.54 & 6 & 14 & 17 \\
M49 & -0.56 & 0.3 & 9.37 & 16.368 & -21.7 & 5.4 & -25.67 & 9 & 11 & 17 \\
N 147 & -2.05 & 0.4 & 10.47 & 0.594 & -13.4 & 7.2 & -16.67 & 6 & 19 & 17 \\
N 205 & -1.40 & 0.15 & 8.92 & 0.752 & -15.46 & 5.59 & -18.79 & 6 & 30 & 17 \\
N 185 & -1.65 & 0.25 & 10.1 & 0.676 & -14.05 & 6.56 & -17.59 & 6 & 23 & 17 \\
M104 & -1.01 & 0.13 & 8.98 & 9.333 & -20.87 & 4.96 & -24.89 & 22 & 11 & 17 \\
N 5128 & -0.96 & 0.04 & 7.84 & 3.8 & -20.06 & 3.94 & -23.96 & 3 & 28 & 17 \\
N 2683 & -1.23 & 0.19 & 10.64 & 7.178 & -18.64 & 6.33 & -22.95 & 27 & 33, 18 & 17 \\
N 7457 & -0.97 & 0.11 & 12.09 & 13.996 & -18.64 & 8.19 & -22.54 & 8 & 11, 7 & 1 \\
N 1023 & -1.29 & 0.11 & 10.35 & 11.117 & -19.88 & 6.24 & -23.99 & 21 & 11 & 17 \\
N 1399 & -0.82 & 0.12 & 10 & 19.055 & -21.4 & 6.31 & -25.09 & 20 & 11 & 17 \\
N 3923 & -0.94 & 0.11 & 10.8 & 22.909 & -21.0 & 6.5 & -25.3 & 24 & 31 & 17 \\
N 524 & -0.97 & 0.12 & 11.3 & 28.2 & -20.95 & 7.16 & -25.09 & 4 & 10 & 1 \\
N 300 & -0.94 & 0.12 & 8.95 & 1.93 & -17.48 & 6.38 & -20.05 & 25, this paper & 15 & 17 \\
\enddata



\tablerefs{(1) 2MASS Extended Source Catalog; (2) Barmby et al. 2000; (3) Beasley et al. 2008; (4) Beasley et al. 2004; (5) Bersier 2000; (6) Brodie and Huchra 1991; (7) Cappellari et al. 2006; (8) Chomiuk, Strader, and Brodie 2008; (9) Cohen, Blakeslee, and Cot\'e 2003; (10) de Vaucouleurs et al. 1991; (11) Ferrarese et al. 2000; (12) Freedman 1990; (13) Freedman et al. 1994; (14) Galleti, Bellazzini, and Ferraro 2004; (15) Gieren et al. 2004; (16) Harris et al. 1996; (17) Jarrett et al. 2003; (18) Jensen et al. 2003; (19) Kang et al. 2007; (20) Kissler-Patig et al. 1998; (21) Larsen and Brodie 2002; (22) Larsen et al. 2002; (23) McConnachie et al. 2004; (24) Norris et al. 2008; (25) Olsen et al. 2004; (26) Perelmuter, Brodie, and Huchra 1995; (27) Proctor et al. 2008; (28) Rejkuba 2004; (29) Schr\"oder et al. 2002; (30) Sharina, Afanasiev, and Puzia 2006; (31) Sikkema et al. 2006; (32) Strader et al. 2003; (33) Tonry et al. 2001.}

\end{deluxetable}

\end{document}